\documentclass[a4paper,12pts]{article}

\usepackage{lipsum}
\usepackage{natbib}
\usepackage{graphics}
\usepackage{amssymb}
\usepackage{amsmath}
\usepackage{graphicx}
\usepackage{amsbsy}
\usepackage{epstopdf}
\usepackage{subcaption}
\usepackage{amsfonts}
\usepackage{setspace}
\usepackage{color}
\usepackage{subfloat}
\usepackage{float}
\usepackage{caption}
\usepackage{pdfpages}
\usepackage{caption}
\usepackage{subcaption}
\usepackage[hidelinks]{hyperref}
\usepackage{algorithm}
\usepackage{pgfgantt}
\usepackage{lscape}
\usepackage{pbox}
\usepackage{tikz}
\usepackage{alltt}
\usepackage{upquote}
\usepackage{float}
\usepackage{array}
\usepackage{pgfplots}
\usetikzlibrary{positioning}
\usepackage{graphicx}
\usepackage{amsmath,mathtools}
\usetikzlibrary{calc,decorations.markings,fit,shapes,chains,arrows,patterns,shadings}

\graphicspath{ {Pictures/} }

\usepackage{titlesec}

\titleformat*{\section}{\large\bfseries}

\usepackage{geometry}
\usepackage[symbol]{footmisc}
\usepackage{authblk}
\usepackage{ragged2e}
\usepackage{enumitem}
\usepackage{tabularx}

\title{Effect of combined heaving and pitching on propulsion of single and tandem flapping foils}
\author[1]{Amit S. Hegde}
\affil[1]{Department of Mechanical Engineering, Birla Institute of Technology \& Science Pilani, Vidya Vihar, Pilani, Rajasthan 333031, India}
\author[2]{Pardha S. Gurugubelli}
\affil[2]{Computing Lab, Department of Mechanical Engineering, Birla Institute of Technology \& Science Pilani, Hyderabad Campus, Hyderabad 500078, India}
\author[3]{Vaibhav Joshi \footnote{Corresponding author: vaibhavj@goa.bits-pilani.ac.in}}
\affil[3]{Department of Mechanical Engineering, Birla Institute of Technology \& Science Pilani, K K Birla Goa Campus, NH 17B Bypass Road, Zuarinagar, Sancoale, Goa 403726, India}
\date{}

\begin{document}

\maketitle

\justifying
\section*{Abstract}
In this study, we present a two and three-dimensional numerical investigation to understand the combined effects of the non-dimensional heave amplitude varying from 0 to 1 and the pitch amplitude ranging from $0^{\circ}$ to $30^{\circ}$ on the propulsive performance for a single and tandem foil system. 
Flow dynamics across single and tandem flapping foils has been considered at Reynolds number of $Re = 1100$ and reduced frequency of $f^* = 0.2$. 
The numerical framework consists of arbitrary Lagrangian-Eulerian moving mesh based algorithm coupled with the variational modeling of the incompressible flow equations. 
%
%The kinematic motion of the foils consists of sinusoidal heaving and pitching motion. 
%
We initially present a systematic analysis on the thrust generation due to the kinematic parameters for a single foil with the aid of effective angle of attack, projected area of the foil to the flow direction, time-averaged pressure and streamwise velocity in the wake \& wake signature. 
%
% The effects have been explained with the help of effective AoA and projected area, time-averaged pressure and streamwise velocity and wake signatures. 
% Questions (related to literature)
%\bluecolor{Such detailed analysis for combined heaving and pitching motion for a single foil has been performed for the first time.}
The significance of effective angle of attack and the projected area of the foil has been emphasized in comprehending the dynamics of lift and drag forces and their relationship with the propulsion.
%and their link with propulsion for a single foil.} 
% Transition from single to tandem
We next investigate the relation between the streamwise gap and kinematic parameters on propulsion for the tandem foil system. We show that the propulsive performance strongly depends on the upstream wake interacting with the downstream foil, and the timing of the interaction due to the gap and phase between the kinematic motion of the foils. 
%and the phase between the type of vortex interaction of the downstream foil with the wake of the upstream foil and its timing during the flapping motion plays an important role in thrust generation. 
% Significance of streamwise gap
Through a control volume analysis for both single and tandem foils, the time-averaged pressure and streamwise velocity have been investigated to explain the effect of kinematic parameters on the hydrodynamic forces.
Typically in the literature, the formation of jet in the wake has been attributed to thrust generation. However, in this study, we emphasize and show the significance of the time-averaged pressure in the wake apart from the streamwise velocity (jet) for predicting the thrust forces.
%in the wake of the foil(s) are crucial indicators to quantify the thrust force. 
The study is concluded with a three-dimensional demonstration of the tandem foils to understand the possible three-dimensional effects due to the large amplitude flapping and wake-foil interaction. %\redcolor{interestingly, it has been observed to two-dimensional flow patterns are observed for the large amplitude flapping and wake-foil interaction of the tandem foil system.}
%null three-dimensionality of the flow is observed at the considered Reynolds number.
%It is found that the effective angle of attack, projected area, type of vortex interaction and the timing of the interaction during the downstroke of the flapping motion influence the propulsive performance. 
%Increasing the heave amplitude of the upstream foil leads to an increase in the average thrust generated by the downstream foil for the gap of 7, while the opposite effect is observed for gap of 4. The propulsion of the downstream foil increases monotonically with the heave amplitude of the downstream foil. Upstream foil's pitch amplitude is noticed to have a minor effect on the performance of the downstream foil, while an increase in the pitch amplitude of the downstream foil increases the thrust for larger gap between the foils. 
%The study of the propulsive performance in the frequency-pitch amplitude parametric space for the tandem foils has been performed for the first time, where a more complex behavior is observed. 
%The trends in the thrust coefficient and the propulsive efficiency are corroborated by the study of various vortex interaction mechanisms leading to favorable or unfavorable thrust generating conditions.

\textbf{Keywords}: Flapping, Tandem, Biomimetics, Pitching, Heaving, Propulsion
	
\section{\label{sec:level1}Introduction}
%--------------------------------------REVISED INTRODUCTION--------------------------------------
Bio-inspiration has been a key factor in the design of underwater vehicles and robots. %Modify the following sentence to incorporate more scientific language 
Recently, there has been an increased interest in learning from nature to improve underwater propulsive performance. The natural propulsion systems utilized by aquatic creatures can out-perform the conventional propulsive devices by as much as 40\%, as mentioned in  \citet{Mannam2020,Yu2005}. Such propulsion systems have many benefits over screw propellers and other traditional propulsors including the absence of cavitation \citep{Huang2013, Arndt2012, Mengjie2020, Mannam2020}, low acoustic signature \citep{Wagenhoffer2021} and excellent manoeuvring performance \citep{Roper2011}, among others.

Bio-inspired underwater propulsion can be broadly classified into two categories \citep{Chu2012}, viz., finned propulsion and jet propulsion. Finned propulsion systems employ flapping foils to generate thrust while jet propulsion utilize flexible membranes and surfaces to squeeze water through a small space (creating a jet). Several species of fish use multiple fins to generate thrust and consequently, some bio-mimetic marine vehicles \citep{Mannam2018, Licht2004_1, Licht2004_2} also use multiple foils for propulsion. Recently, flapping foils have been employed in marine engineering to harvest wave energy in \citet{JI2017185}, power generation by turbines in \citet{Kinsey2012}, ship propulsion in  \citet{BELIBASSAKIS2013227} and in propulsion of unmanned underwater vehicles (UUVs) \citep{Ramamurti2010, ZHANG2021108763}. The flapping foil system has very high potential to enhance propulsion and stability of ships and UUVs \citep{1353412}. If the fins/foils are arranged in-line, the configuration is known as a \textit{tandem} configuration. On the other hand, if the foils are arranged such that they are parallel to one another, the configuration is said to be \textit{side-by-side}. 

In addition to making use of flapping fins/foils, fish often swim together in formations known as ``schools''. It has been shown that this collective swimming behavior offers a hydrodynamic advantage to a fish within the formation as a result of the wake created by fish leading the school \citep{Marras2012}. The enhanced hydrodynamic performance corresponds to the tandem configuration of fins/foils and could be beneficial for underwater vehicles. To create robust and reliable tandem flapping foil propulsors, it is essential that the physics behind the relevant flow phenomena is well understood. Therefore, in this study we focus on the flapping dynamics of a tandem foil configuration. 

Literature pertaining to flapping foils highlight several governing parameters such as Reynolds number of the flow, foil geometry (chord length and thickness), and foil kinematics \citep{WU2020106712}. In addition, the frequency of flapping is also of great significance. Based on the combinations of above-mentioned parameters, a flapping foil can either produce thrust or extract energy from the surrounding flow, as noted by \citet{kinsey2006}. Several studies have been carried out in the energy extraction regime for single as well as tandem flapping foils by \citet{Xu2019, Karbasian2015, Ma2021, Ribeiro2021}. In the case of tandem foils, the phase difference in the flapping and the streamwise gap between the foils are also crucial parameters \citep{WU2020106712}.

Several works have investigated the relationship between the phase difference of tandem flapping foils and the propulsive performance. As studied by \citet{Cong2020}, the performance of the downstream foil is affected significantly by the phase difference between the upstream and the downstream foil for streamwise gap distances of 0.25-0.75 chord lengths at $Re = 200$. Studies carried out by \citet{Sampath2020} at $Re = 36500$ have revealed that the downstream foil generates more thrust than the upstream foil when it lags by a quarter cycle, but performs worse if it leads by the same amount. It has also been found that the downstream foil generates maximum thrust when the flapping of both the foils is in-phase at $Re = 5000$ by \citet{Lua_2016}. Propulsive performance of in-line multiple foils with fixed spacing of $g/c = 0.25$ at Reynolds numbers of $500$ and $1000$ were considered by \citet{HAN2022103422} where the effects of phase difference and Reynolds number were investigated.

Apart from the phase difference, the gap between the tandem foils has significant effect on the flow dynamics and propulsive performance. For synchronized plunging foils, \citet{Chen2022} found that at $Re = 5000$, thrust enhancement is maximum when the streamwise gap between the foils is between 1.5 and 2 chord lengths. 
%The use of finite-thickness flat plates in place of foils is not uncommon in the realm of two-dimensional analysis. 
Studies on tandem self-propelled flexible flapping plates have also been carried out by \citet{Ryu2020, Peng2018}. At Reynolds number of 100, significant improvements in thrust generation was obtained by reducing the streamwise gap by \citet{Ryu2020}. At $Re = 200$, the propulsive efficiency was found to be larger when the upstream plate was longer than the downstream plate \citep{Peng2018}. A numerical study by \citet{Pan2020} at $Re = 1000$ identified that an increase in the streamwise spacing in a school of flapping foils reduced the influence of the lateral neighbors on the performance of the flapping foils. Recently, \citet{OMAE_Joshi_2021, Joshi2021} studied the effect of gap and the chord sizes for the tandem foils on the propulsion at $Re = 1100$. The mechanism of wake interaction with the downstream foil was identified and studied in detail. A periodic variation in the thrust performance for the downstream foil was observed with the gap between the foils, indicating the crucial effect of the wake interaction. Furthermore, the combined effects of phase difference and the gap between the tandem foils was studied experimentally by \citet{Boschitsch2014}.
%The effect of oscillation frequency has received significant attention in the context of a single flapping foil. Several numerical simulations by \cite{Das2016, Deng_2016, Yu2017} have examined the effect of flapping frequency on the propulsive performance. \cite{Gupta2021} numerically examined the effect of Strouhal number $St$ and Reynolds number $Re$ on the propulsive performance of a National Advisory Committee for Aeronautics (NACA) 0012 foil. It was found that the time averaged thrust coefficient exhibits a quadratic relationship with increasing $St$ and is greatly enhanced by $Re$. 
%\cite{Gungor2020} investigated the effect of wake symmetry on the performance of a tandem hydrofoil configuration. Both in-phase and out-of-phase oscillations were considered at high $St$ and found to exhibit unsteady behavior with distinct wake characteristics. The influence of Strouhal number and stream-wise gap for a tandem foil configuration has also received some attention in the literature [\cite{Simsek2020, WANG2021106939, BROERING2015124, Gravish2015}]. 
The gap between the tandem foils were considered from 0.5 to 5 times the chord length along with varying Strouhal numbers and phase difference by  \citet{muscutt_weymouth_ganapathisubramani_2017} at Reynolds number of 7000. It was found that the downstream foil produced from null to almost twice the thrust of a single foil, depending on the gap and phase difference. The vortex interaction between the foils were also discussed in detail.

Works by \citet{Mandujano2018, Alam2020, Chao2021_2, Chao2021_1, Thakor2020, Das2016} dealt with pure pitching motion of the foil. Non-dimensionalizing with respect to the foil thickness $d$, \citet{Alam2020} studied pitching foil at $0.21 \geq St_d = fd/U_{\infty} \geq 0.33$, $1.1 \geq A^* = A/d \geq 1.6$, and \citet{Chao2021_2, Chao2021_1} considered $St_d = 0.1-0.3$ and $A^* = 0.5-2$, where $A$ denotes the peak-to-peak amplitude of the trailing edge. Furthermore, the parameters utilized by \citet{Thakor2020} consisted of $2^{\circ}-6^{\circ}$ of pitch amplitude and reduced frequency $(\pi f c/U_{\infty})$ in the range $3 - 9$, $c$ being the chord of the foil. Studies by \citet{Deng_2016, floryan2017} focused on pure pitching and heaving motions separately. %and developed scaling laws for the thrust and power coefficients. 
%They selected pitch amplitude in the range $3^{\circ} - 15^{\circ}$, dimensionless heave amplitude $h_0/c = 0.06 - 0.19$ and $St_A = fA/U_{\infty} = 0.05 - 4$, where the Strouhal number is defined on the basis of amplitude of trailing edge.
Experiments conducted by \citet{Floryan2019} involved combined heaving and pitching motion where $h_0/c = 0.1 - 0.75$, pitch amplitude $\theta_0 = 5^{\circ} - 40^{\circ}$ with frequency of 0.1 Hz and chord-based Reynolds number of 8000. Computations by \citet{Yu2017} considered frequency range $1 - 30$ Hz, pitch amplitude $3^{\circ} - 19^{\circ}$, $h_0/c = 0.1 - 0.9$ and Reynolds number $1000 - 1600000$.
%deal with the effect of these parameters on a single foil. The Strouhal number - pitch amplitude parametric space was studied for a single foil and a rich spectrum of wake behavior for pure pitching motion was observed by \citet{Chao2021_2, Chao2021_1}. 
%In the case of asymmetric pitching foil motion by  \citet{Thakor2020}, the mean thrust coefficient was noted to increase with pitch amplitude at $Re = 3000$ and $Re = 10000$. 
%The effect of heave amplitude was considered by \citet{floryan2017} where its importance at low reduced frequency was noted. 
As per the knowledge of the authors, studies dealing with the combined motion of heaving and pitching of the single flapping foil are scarce and no such detailed study concerning the kinematic motion parameters exists for the tandem foil configuration. 
%Therefore, combined heaving and pitching motion of single and tandem foils at large heave and pitch amplitudes have not been comprehensively investigated.

Most of the works in the literature talk about the transition from the drag-producing  von-K$\mathrm{\acute{a}}$rm$\mathrm{\acute{a}}$n vortex street to the thrust-producing inverted von-K$\mathrm{\acute{a}}$rm$\mathrm{\acute{a}}$n street in the wake due to change of kinematic parameters in the flapping motion of the foil. 
%This transition indicates the , while comparing the effects of kinematic motion on propulsion, change of von-K$\mathrm{\acute{a}}$rm$\mathrm{\acute{a}}$n vortex street to inverted von-K$\mathrm{\acute{a}}$rm$\mathrm{\acute{a}}$n street in the wake of the foil indicates the transformation from drag-producing to thrust-producing regimes. 
% TRANSITION (1)
It has also been noted that while investigating the thrust-producing regime, researchers have mostly relied on the time-averaged streamwise velocity in the wake of the foil to comprehend the thrust generation. %The formation of jet in the wake instead of a velocity-deficit region indicates a thrust-producing scenario. 
It was pointed out in \citet{Alam2020} that the jet formation and wake signature are attributes of the thrust production and do not give an insight about its origin. The trend of mean thrust generation still remains unanswered for the scenarios where flapping foils with positive thrust are compared. Is the comparison of time-averaged streamwise velocity in the wake for these cases enough to predict the trend in the mean thrust force? We try to answer this question in the current work by considering the influence of the kinematic motion parameters (heave and pitch amplitudes) on propulsion for single as well as tandem foils.

Majority of the computational research conducted on flapping foils has been two-dimensional \citep{WU2020106712} and the three-dimensional spanwise as well as end effects have not been taken into consideration. Recent works by \citet{Lagopoulos2021, Arranz2020, Jurado2022} have 
%focused on finite foils as well as end effects and 
studied the effect of aspect ratio of the foil on propulsive performance. However, the conditions under which three-dimensional (3D) flow effects become important are yet to be studied in detail for tandem foils.  

%The Reynolds number range $10^3 - 10^4$ has been found to be relevant for bird flight and swimming of fishes \citep{Taylor2003, TRIANTAFYLLOU1993205}. Studies carried out by \citet{ASHRAF2011145} concluded that cambered foils offer no propulsive advantage compared to symmetrical foils. Furthermore, at low Reynolds numbers, thin foils were observed to be better than thick foils due to the weakening of the peak of the suction pressure as a consequence of more rounded leading edge of thicker foils. Therefore, the current work deals with the low Reynolds number regime pertaining to biological locomotion along with thin foils as a two-dimensional assumption of the wing/fin.

%To the best of the authors' knowledge, the effects of heave and pitch amplitudes of the two foils in the tandem configuration on their propulsive performance as a combined flapping (heaving + pitching) motion have not been explored in detail. 
%Although the effect of reduced frequency and $St$ have been studied in detail for a single flapping foil, the literature lacks in the study of these parameters for the tandem foil configuration. Moreover, the impact of the collective frequency-pitch amplitude parametric space on tandem foil propulsion can give an insight about possible optimal operational regimes of flapping foil propulsion systems.   
In the present study, we numerically investigate the flow dynamics of a single and tandem flapping foil system at low Reynolds number of 1100. This Reynolds number falls in the range $10^3 - 10^4$ which has been found to be relevant for bird flight and swimming of fishes \citep{Taylor2003, TRIANTAFYLLOU1993205}. We employ a moving mesh arbitrary Lagrangian-Eulerian framework for the flapping motion of the foil. The fluid dynamics is modeled with the help of variational finite element method applied to incompressible Navier-Stokes equations.

We try to shed some light on the following questions from the present work:

$\bullet$ How does the kinematic parameters such as heave and pitch amplitudes affect the propulsive performance for combined heaving and pitching of single and tandem foils?

$\bullet$ Can the trend in mean thrust force be predicted solely by observing the time-averaged streamwise velocity in the wake of the flapping foil (single and tandem) system?

$\bullet$ What is the influence of the streamwise gap between the tandem foils on the thrust generation capability of the downstream foil?

$\bullet$ How does the three-dimensional flow behave during the wake-foil interaction for large amplitude flapping foils in tandem configuration?

%The salient and novel contributions from the present work are as follows:

%$\bullet$ Effect of combined heave and pitch amplitudes on the thrust performance of single and tandem flapping foil(s),

%$\bullet$ Investigation of the hydrodynamic performance and its relationship with the effective angle of attack and projected area with the incoming flow for a single foil,

%$\bullet$ Discussion about the variation of mean thrust as a consequence of both mean pressure and streamwise velocity distribution in the wake of the single/tandem foil(s) via a control volume approach,

%$\bullet$ Comparison of the effect of the kinematic parameters (heave and pitch amplitudes) on propulsion of the tandem system for different streamwise gap between the foils,

%$\bullet$ Comprehensive insights about the wake-vortex interactions and wake signature of the flapping foil(s) and their link with propulsive performance, and

%$\bullet$ The three-dimensional spanwise behavior of the flow around the tandem foils \redcolor{at $Re = 1100$ for large amplitude flapping}.

% Layout of the present paper
The article is organized in the following manner. First, we briefly discuss the numerical framework in section \ref{numerical_framework}. The next section \ref{poi} discusses the definition of the various parameters utilized in the study. Flapping dynamics of a single foil along with the effects of the kinematic parameters are studied in section \ref{flapping_single}. The tandem arrangement of flapping foils is examined in section \ref{tandem_flapping_foil}. 
This is followed by the demonstration of the three-dimensional simulation for tandem foils in Section \ref{3d_demo}. 
Finally, the key findings are summarized and the study is concluded in section \ref{conclusion}.

\section{Numerical framework}
\label{numerical_framework}

In the current study, the flapping dynamics of the foils is modeled using the moving mesh arbitrary Lagrangian-Eulerian (ALE) framework. Discretization of the flow equations is performed using a stabilized Petrov-Galerkin variational formulation, while the foil motion is specified by satisfying the kinematic equilibrium condition or the velocity continuity at the interface between the fluid and the foil. Here, we briefly review the governing equations of the formulation for the sake of completeness.

%%%%%%%%%%%%%%%%%%%%%%%%%%%%%%%%%%%%%%%%%%%%%%%%%%%%%%%%%%%%%%%%%%%%%%
The flow is modeled with the help of incompressible Navier-Stokes equations written in the ALE framework as
\begin{align} \label{NS_1}
	\rho^{\rm{f}} \frac{\partial \boldsymbol{v}^{\rm{f}}}{\partial t}\bigg|_{\boldsymbol{\chi}} + \rho^{\rm{f}}(\boldsymbol{v}^{\rm{f}} - \boldsymbol{w})\cdot\nabla\boldsymbol{v}^{\rm{f}} &= \nabla\cdot\boldsymbol{\sigma}^{\rm{f}} + \rho^{\rm{f}}\boldsymbol{b}^{\rm{f}},\\
	\nabla\cdot\boldsymbol{v}^{\rm{f}} &= 0, \label{NS_2}
\end{align}
where the fluid velocity is given by $\boldsymbol{v}^{\rm{f}} = (v^\mathrm{f}_x, v^\mathrm{f}_y)$ with its X- and Y- components, and the mesh velocity is denoted as $\boldsymbol{w}$. The body force acting on a fluid element is written as $\boldsymbol{b}^{\rm{f}}$ and the fluid density is given as $\rho^{\rm{f}}$. We consider a Newtonian fluid for which the Cauchy stress tensor can be written as $\boldsymbol{\sigma}^{\rm{f}} = -p\boldsymbol{I} + \mu^{\rm{f}}(\nabla\boldsymbol{v}^{\rm{f}} + (\nabla\boldsymbol{v}^{\rm{f}})^T)$ in which the fluid pressure is denoted by $p$, fluid dynamic viscosity by $\mu^{\rm{f}}$ and the identity matrix by $\boldsymbol{I}$. In Eq. (\ref{NS_1}), $\boldsymbol{\chi}$ denotes the ALE referential coordinate system pertaining to the moving mesh coordinates. The governing equations are temporally discretized in the time interval $t \in [t^{\rm{n}}, t^{\rm{n+1}}]$ with the help of the Generalized-$\alpha$ method by \citet{Gen_alpha} while the spatial discretization is carried out by stabilized finite element approximations. Detailed description of the present formulation can be found in the works by  \citet{JOSHI2018137, Joshi_IJNME_2019, FSI_Book}.

% Fig 1
\begin{figure}
	\centering
		\includegraphics[width=\textwidth]{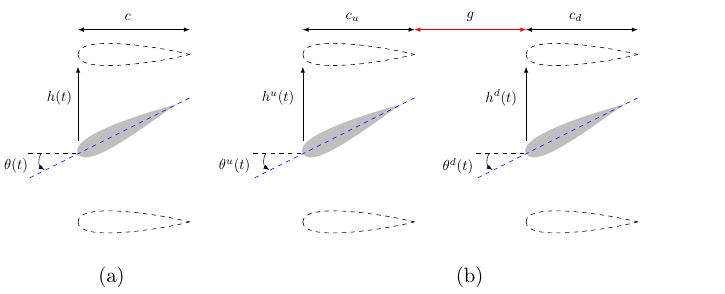}
\caption{Flapping kinematic motion of the foil consisting of heaving and pitching motion for (a) single foil, and (b) tandem foils.}
	\label{foil_kinematics}
\end{figure}

The flapping kinematic motion of the foil is illustrated in Fig. \ref{foil_kinematics}. For a single foil as shown in Fig. \ref{foil_kinematics}(a), the kinematics consists of a heave component $h(t) = h_0\mathrm{sin}(2\pi f t + \phi_h)$ along with a pitch component $\theta(t)=\theta_0 \mathrm{sin}(2\pi f t)$ about a pitching axis located at the leading edge of the foil. Here, $h_0$, $\theta_0$, $f$ and $\phi_h$ denote the heave amplitude, pitch amplitude, flapping frequency and the phase difference between the heaving and pitching motion, respectively. We also consider the tandem configuration of flapping foils with a gap of $g$ between the foils (Fig. \ref{foil_kinematics}(b)). The chord length of the upstream and the downstream foils are denoted by $c_u$ and $c_d$, respectively. The upstream foil's motion is given as
\begin{align}
    \theta^u(t) &= \theta_0^u \mathrm{sin}(2\pi f^u t),\\
    h^u(t) &= h_0^u\mathrm{sin}(2\pi f^u t + \phi_h^u),
\end{align}
where $h_0^u$, $\theta_0^u$, $f^u$ and $\phi^u_h$ represent the heave amplitude, pitch amplitude, flapping frequency and phase difference between the heaving and pitching motion, respectively, for the upstream foil. Similarly, the motion of the downstream foil is given by
\begin{align}
    \theta^d(t) &= \theta_0^d \mathrm{sin}(2\pi f^d t + \varphi),\\
    h^d(t) &= h_0^d\mathrm{sin}(2\pi f^d t + \phi_h^d + \varphi),
\end{align}
where $\varphi$ is the phase difference between the kinematic motions of the upstream and downstream foils and the other symbols have their usual meanings. In the current work, we consider National Advisory Committee for Aeronautics (NACA) 0015 foils in tandem with $c_u/c_d = 1$, $\varphi = 0^{\circ}$ and $\phi_h^u = \phi_h^d = 90^{\circ}$. 

The current problem involves an interaction between the fluid and structure in which the structural displacements are imposed on the surface of the foil. This means that there is a one-way coupling that is facilitated by matching the structural and fluid velocities at the boundary between the foil and the fluid (kinematic equilibrium condition). The mathematical equation associated with this boundary condition is given as
\begin{align}\label{kinematic_equilibrium}
	\boldsymbol{v}^{\rm{f}} (\widetilde{\boldsymbol{\varphi}}(\boldsymbol{X},t),t) &= \boldsymbol{v}^{\rm{s}} (\boldsymbol{X},t),\ \forall \boldsymbol{X} \in \Gamma^\mathrm{fs},
\end{align}
where $\widetilde{\boldsymbol{\varphi}}$ represents a one-to-one mapping between the structural position $\boldsymbol{X}$ at time $t=0$ and its corresponding position at time $t > 0$. $\Gamma^{\mathrm{fs}}$ is the fluid-structure interface at $t = 0$ and $\boldsymbol{v}^\mathrm{s}$ denotes the structural velocity.

The nonlinear Navier-Stokes equations are solved by the Newton-Raphson iterative technique. The coupling between the flow equations and the moving mesh framework is carried out in a partitioned iterative manner, the details of which can be found in the work by \citet{Joshi2021, FSI_Book}. The above formulation has been verified and validated, consisting of mesh convergence and time convergence studies in the earlier work by \citet{Joshi2021} for the single and tandem flapping foils and will not be discussed in the present work for brevity.

\section{Parameters of interest}
\label{poi}
The non-dimensional parameters pertaining to the single flapping foil are the Reynolds number $Re = (\rho^\mathrm{f} U_{\infty}c)/\mu^{\mathrm{f}}$, non-dimensional heave amplitude $h_0/c$ and non-dimensional flapping frequency $f^* = (fc)/U_{\infty}$, where $U_{\infty}$ is the freestream velocity of the flow. Similarly, for the tandem foils, we consider the characteristic length as the chord of the downstream foil $c_d$. Thus, Reynolds number $Re = (\rho^\mathrm{f} U_{\infty}c_d)/\mu^{\mathrm{f}}$, non-dimensional heave amplitude of upstream and downstream foils are denoted by $h^u_0/c_d$ and $h^d_0/c_d$, non-dimensional flapping frequency of upstream and downstream foils are given by $f^*_u = (f^uc_d)/U_{\infty}$ and $f^*_d = (f^dc_d)/U_{\infty}$, respectively, and the gap ratio between the foils is $g/c_d$. Note that we consider $c_u/c_d = 1$ in the present study.

The propulsive performance of the flapping foils is determined by evaluating the integrated values of the fluid forces on the foil surface. The instantaneous coefficients are given as follows for a single foil:
\begin{align}
	C_Y &= \frac{F_Y}{\frac{1}{2}\rho^\mathrm{f}U_{\infty}^2 c l} = \frac{1}{\frac{1}{2}\rho^\mathrm{f}U_{\infty}^2 c l} \int_{\Gamma^\mathrm{fs}(t)} (\boldsymbol{\sigma}^\mathrm{f} \cdot \boldsymbol{n})\cdot \boldsymbol{n}_y d\Gamma, \label{CL_def}\\
	C_X &= \frac{F_X}{\frac{1}{2}\rho^\mathrm{f}U_{\infty}^2 c l} = \frac{1}{\frac{1}{2}\rho^\mathrm{f}U_{\infty}^2 c l} \int_{\Gamma^\mathrm{fs}(t)} (\boldsymbol{\sigma}^\mathrm{f} \cdot \boldsymbol{n})\cdot \boldsymbol{n}_x d\Gamma,\\
	C_T &= \frac{-F_X}{\frac{1}{2}\rho^\mathrm{f}U_{\infty}^2 c l} = -C_X,\\
	C_P &=\frac{P}{\frac{1}{2}\rho^\mathrm{f}U_{\infty}^3 cl} =  \frac{-F_Y v^\mathrm{s}_{y,\mathrm{heave}} - M_Z \omega }{\frac{1}{2}\rho^\mathrm{f}U_{\infty}^3 cl}. \label{CP_def}
\end{align}
In the set of equations given above, $C_Y$ and $C_X$ are the force coefficients in the transverse and inline directions to the freestream velocity $U_{\infty}$ respectively, $C_T$ is the thrust coefficient and $C_P$ denotes the power coefficient. $F_X$, $F_Y$ and $P$ are the X-component of the force, Y-component of the force and the power supplied to the structure respectively. The moment of the forces acting about the pitching axis of the foil is denoted by $M_Z$ while the heave translational velocity of the foil is written as $v^\mathrm{s}_{y,\mathrm{heave}} = 2\pi f h_0 \mathrm{cos}(2\pi f t + \phi_h)$. The angular velocity of the foil in pitching motion is denoted by $\omega$. Here, $c$ and $l = 1$ are the chord and the span of the foil respectively. The variable $\overline{X}$ represents the time averaged mean value of $X$ over a time period $T$ of the oscillation. The propulsive efficiency of the foil can thus be written as $\eta = \overline{C_{T}} / \overline{C_{P}}$. Similarly, one can extend the coefficients for the tandem foils. The mean thrust coefficient and propulsive efficiency of the combined tandem foil system can be evaluated by considering the combined mean thrust and power coefficients of the foils.

Next, we discuss the propulsive performance of a single and tandem foil configurations subjected to a freestream flow under the variation of the heave amplitude and pitch amplitude. We investigate the different mechanisms of thrust generation to comprehensively understand the flow dynamics of flapping foils for such scenarios.

%====================================================================================
\section{Flapping of a single foil}
\label{flapping_single}

The single flapping foil system has received significant attention in the context of both numerical and experimental studies. In particular, studies on the effect of kinematic parameters on the propulsive performance  of the foil have been performed by \citet{Das2016, Deng_2016, Yu2017, Mandujano2018, Alam2020, Chao2021_2, Chao2021_1, Thakor2020, floryan2017} for a range of Reynolds number, flapping frequency, heave and pitch amplitudes. However, the combined motion of heaving and pitching for a single foil and the influence of the kinematic parameters on propulsion have not been studied comprehensively and no such detailed study exists for the tandem foils. Prior to understanding the effects of heave and pitch amplitudes on the performance of the tandem foil configuration, we discuss their effects for a single isolated foil in this section. 

We perform two-dimensional computations and give insights about the influence of varying heave and pitch amplitudes and explain the trends with the help of wake signatures. %We review the flow across a single flapping foil and the effect of kinematic parameters which are different in comparison to the aforementioned studies. This has been carried out for systematic comparison of the tandem foil system with the isolated single foil. 
Furthermore, we comprehend the generation of thrust with the help of a control volume analysis which gives a complete picture of the time-averaged propulsive performance for the foil.

%Floryan et al. \cite{floryan2017} conducted an experimental study involving a heaving and pitching foil. The pitch amplitude was varied from 3 to 15$^\circ$ in intervals of $2^\circ$ while the heave amplitude spanned across 5 to 15 mm with intervals of 2 mm. The amplitude based Strouhal number was kept between 0.05 and 0.4 with a step change of 0.025. For these set of parameters, it was found that at $Re = 4780$, pitching motion produced maximum thrust at moderate values of St. In the present work, a similar behavior is observed for a single flapping foil. It is found that the propulsive efficiency peaks at a dimensionless frequency between 0.15 and 0.20 (Fig. \ref{CTmean_Eta_f_theta0_single}). Das et al. \cite{Das2016} conducted viscous vortex particle simulations of a single pitching foil for amplitudes between 2 and 16 $^\circ$ and $Re\in[10,2000]$. Deng et al. \cite{Deng_2016} performed a similar study at higher values of $Re\in[10,1600000]$ and at a pitch amplitude of 8$^\circ$. Both studies reported that the maximum propulsive efficiency for this range of $Re$ and pitch amplitude is approximately 16\%. In comparison, the present study indicates that at 8$^\circ$, the maximum propulsive efficiency is around 15\% (Fig. \ref{CTmean_Eta_f_theta0_single}) which is in good agreement with the value reported in existing literature.}

\begin{figure}
	\centering
	\includegraphics[width=\textwidth]{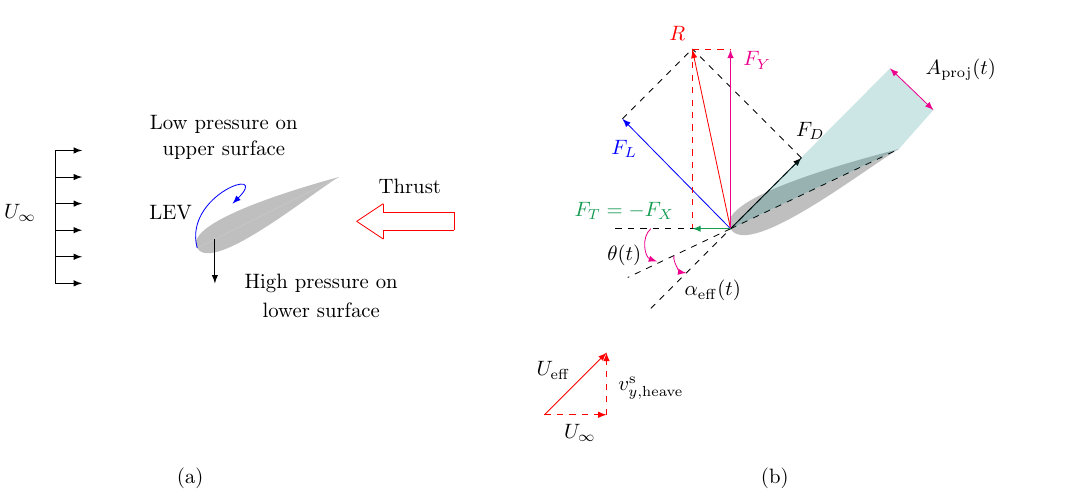}
\caption{Flapping single foil: (a) generation of thrust as a result of LEV during the downstroke, and (b) instantaneous force components during flapping.}
	\label{LEV_single_schematic}
\end{figure}
A flapping foil generates thrust through the development of a leading edge vortex (LEV) in the propulsion regime. During the downstroke, this LEV is responsible for suction pressure (negative pressure) on the upper surface of the foil, whereas a positive pressure exists at the lower surface. This pressure differential along with the orientation of the flapping foil during the flapping motion leads to a net thrust force. This is depicted in Fig. \ref{LEV_single_schematic}(a). Therefore, the favorable conditions for generation of thrust during the downstroke of a flapping foil are \citep{Joshi2021}: (i) suction pressure on the upper surface, and (ii) positive pressure on the lower surface. 

The various components of the net resultant force $R$ on the foil are shown in Fig. \ref{LEV_single_schematic}(b). As a result of the  incoming freestream velocity and the heave velocity of the foil, the incoming flow effective velocity $U_\mathrm{eff}$ is inclined to the horizontal direction by an angle $\mathrm{tan}^{-1} (-v^\mathrm{s}_{y,\mathrm{heave}}(t)/U_{\infty})$. The effective angle of attack for the foil can thus be defined as 
\begin{align}
    \alpha_{\mathrm{eff}}(t) &= \mathrm{tan}^{-1} \bigg( -\frac{v^\mathrm{s}_{y,\mathrm{heave}}(t)}{U_{\infty}}  \bigg) - \theta(t),
\end{align}
which represents the inclination of $U_\mathrm{eff}$ with the chord of the flapping foil. The components of the force along the effective velocity and perpendicular to it are known as the drag $(F_D)$ and lift $(F_L)$ forces, respectively. The resultant force can also be decomposed in the direction of freestream velocity as $F_X$ and perpendicular to it as $F_Y$. A relationship exists between these two decompositions and can be written as
\begin{align}
    F_D &= F_X~ \mathrm{cos}(\alpha_\mathrm{eff} + \theta) + F_Y~ \mathrm{sin}(\alpha_\mathrm{eff} + \theta),\\
    F_L &= -F_X~ \mathrm{sin}(\alpha_\mathrm{eff} + \theta) + F_Y~ \mathrm{cos}(\alpha_\mathrm{eff} + \theta).
\end{align}
The drag $(C_D)$ and lift $(C_L)$ coefficients can also be defined by non-dimensionalizing these forces with $(1/2)\rho^\mathrm{f}U_{\infty}^2 cl$.

We also quantify the projected area of the foil as seen from the effective flow direction $U_\mathrm{eff}$. For this scenario, the foil is assumed to be a straight line connecting the leading edge to the trailing edge. Based on the prescribed flapping motion, the projected area is evaluated as
\begin{align}
    A_{\mathrm{proj}}(t) &= | c~\mathrm{sin}(\alpha_\mathrm{eff}(t)) |l,
\end{align}
where $l=1$ is the span of the foil.

\subsection{Effect of heave amplitude $(h_0)$ on propulsion}
\label{single_h0_effects}

\begin{figure}
		\centering
		\includegraphics[width=\textwidth]{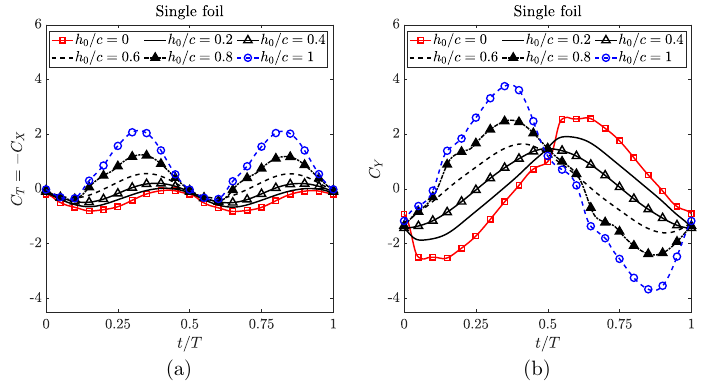}
        \caption{Temporal variation of the force coefficients in a flapping cycle for a single foil with varying $h_0/c$ (fixed $\theta_0 = 30^{\circ}$): (a) $C_T$, and (b) $C_Y$.}
        \label{TH_single_h0_CtCy}
\end{figure}

\begin{figure}
		\centering
		\includegraphics[width=\textwidth]{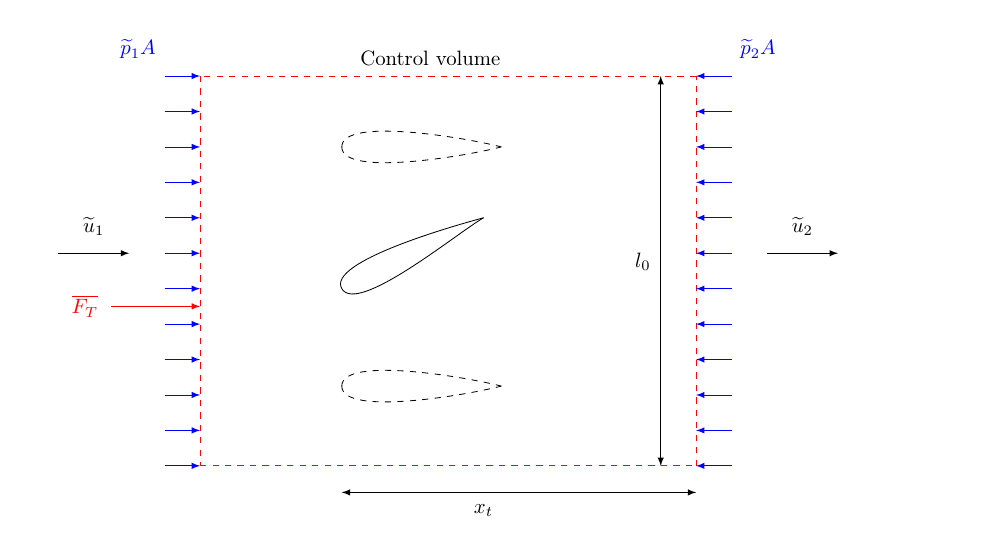}
        \caption{Control volume analysis of the flapping foil.}
        \label{Single_CV}
\end{figure}

The temporal variation of the force coefficients in the X and Y directions in a flapping cycle considering $f^* = 0.2$, $\theta_0 = 30^{\circ}$ and $Re=1100$ is shown for different heave amplitudes in Fig. \ref{TH_single_h0_CtCy}. It can be observed that an increase in the heave amplitude leads to higher thrust generation for the single foil with maximum thrust coefficient noted for $h_0/c = 1$.
It can be deduced that the mean thrust coefficient $\overline{C_T}$ over a flapping cycle increases with increase in heave amplitude of the foil, where $\overline{C_T}$ is negative for $h_0/c = 0$ and transitions to a positive thrust as heave amplitude increases. To understand this average variation of the thrust with heave amplitude, we consider a control volume surrounding the flapping foil, as shown in Fig. \ref{Single_CV}. We will apply the conservation of linear momentum principle in the streamwise direction. Let $\widetilde{p}_1 A$ and $\widetilde{p}_2 A$ denote the pressure forces on the left and right boundaries of the control volume, $\widetilde{u}_1$ and $\widetilde{u}_2$ represent the velocities of the fluid at the boundaries and $\overline{F_T}$ is the reaction to the thrust force on the control volume. 
\begin{figure}
		\centering
		\includegraphics[width=\textwidth]{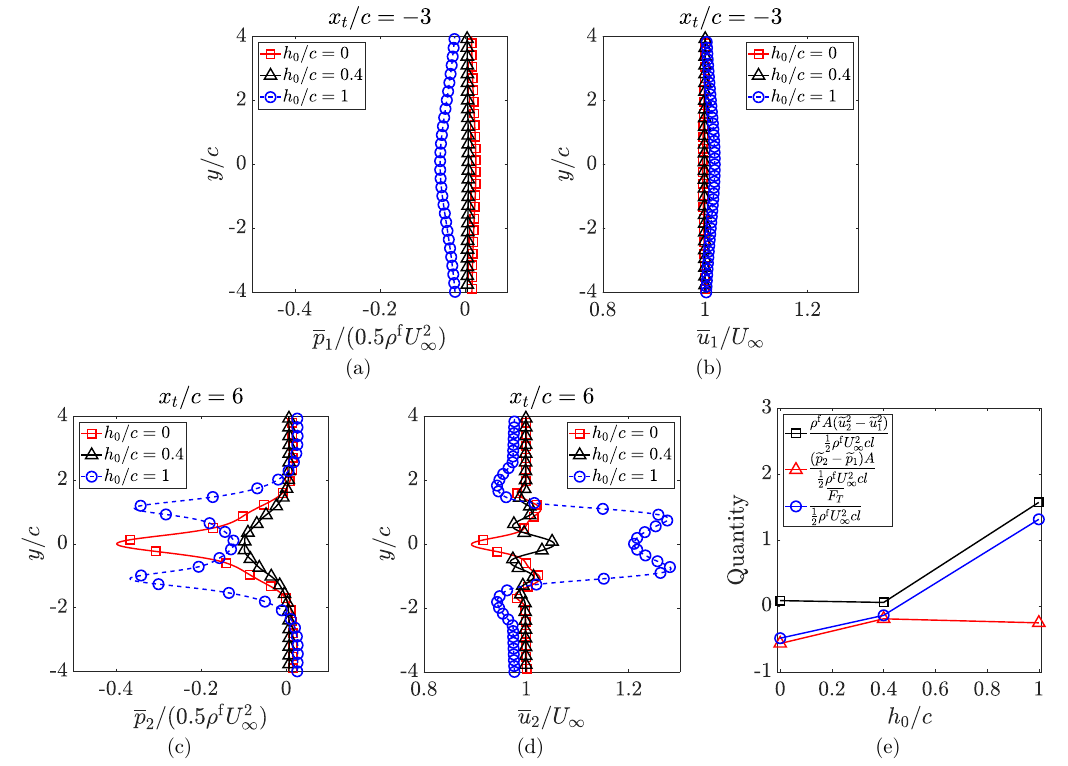}
        \caption{Control volume analysis of the flapping foil for the effect of heave amplitude: (a) time-averaged pressure at the left boundary $(\overline{p}_1)$, (b) time-averaged streamwise velocity at the left boundary $(\overline{u}_1)$, (c) time-averaged pressure at the right boundary $(\overline{p}_2)$, (d) time-averaged streamwise velocity at the right boundary $(\overline{u}_2)$, and (e) evaluated quantities based on Eq. (\ref{Mom_Bal_Eq}) for varying $h_0/c$.}
        \label{Single_h0_CV_quantity}
\end{figure}
Following Newton's second law, 
\begin{align}
    \displaystyle\sum F_x &= \dot{m}_2 \widetilde{u}_2 - \dot{m}_1 \widetilde{u}_1 \\
    \widetilde{p}_1 A - \widetilde{p}_2 A + \overline{F_T} &= \rho^\mathrm{f} A (\widetilde{u}_2^2 - \widetilde{u}_1^2 ) \\
    \overline{F_T} &= \rho^\mathrm{f} A (\widetilde{u}_2^2 - \widetilde{u}_1^2 ) + (\widetilde{p}_2 - \widetilde{p}_1) A.\label{Mom_Bal_Eq}
\end{align}
As can be observed, the mean thrust force on the foil is a consequence of not just the change in velocity but also the change in the pressure across the control volume. Note here that the qualtities $\widetilde{p}_1$, $\widetilde{p}_2$, $\widetilde{u}_1$ and $\widetilde{u}_2$ are evaluated in an averaged sense and vary uniformly over the two boundaries. They are defined as follows:
\begin{align}
    \widetilde{p}_1 A &= \int_0^{l_0} \overline{p}_1 l~dy, \label{int_p1}\\
    \widetilde{u}_1^2 A &= \int_0^{l_0} \overline{u}_1^2 l~dy, \label{int_u1}
\end{align}
where $l_0$ and $l$ are the width of the control volume and the span (in the third dimension, $l = 1$ for two-dimensional study), respectively. Here, $\overline{p}_1$ and $\overline{u}_1$ are the time-averaged variation of the quantities across the Y-direction, as shown in Figs. \ref{Single_h0_CV_quantity}(a) and (b), respectively. Similar expressions can be defined for $\widetilde{p}_2 A$ and $\widetilde{u}_2^2 A$ for which the time-averaged quantities across Y-direction are also depicted in Figs. \ref{Single_h0_CV_quantity}(c) and (d).

For the current scenario of varying heave amplitude, it can be noticed from Figs. \ref{Single_h0_CV_quantity}(a) and (b) that the variation in $\overline{u}_1$ and $\overline{p}_1$ across different $h_0/c$ is negligible as they represent freestream quantities. The difference between the cases is observed for $\overline{u}_2$ and $\overline{p}_2$. The peak of the time-averaged velocity $\overline{u}_2$ is negative for $h_0/c = 0$ depicting a velocity-deficit region in the wake, while it transitions to a jet-like velocity profile for $h_0/c = 1$ (see Fig. \ref{Single_h0_CV_quantity}(d)). On the other hand, the negative pressure is larger for $h_0/c = 0$ compared to $h_0/c = 0.4$ (Fig. \ref{Single_h0_CV_quantity}(c)). For $h_0/c = 0$, the velocity is small, while the pressure has a large negative value, resulting in a larger contribution of the $(\widetilde{p}_2 - \widetilde{p}_1) A$  term in Eq. (\ref{Mom_Bal_Eq}) giving negative thrust, which is clearly visible in Fig. \ref{Single_h0_CV_quantity}(e). As the heave amplitude is increased to $h_0/c = 1$, the velocity attains large values in the wake making the contribution of $\rho^\mathrm{f} A (\widetilde{u}_2^2 - \widetilde{u}_1^2)$ term larger resulting in higher thrust. Thus, the increasing trend of the mean thrust coefficient can be explained by observing the variations in both velocity and pressure at the downstream wake of the flapping foil as summarized in Fig. \ref{Single_h0_CV_quantity}(e).

\begin{figure}
		\centering
		\includegraphics[width=\textwidth]{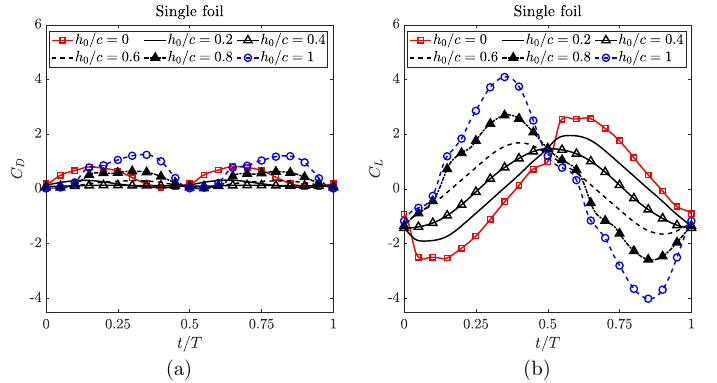}
        \caption{Temporal variation of the force coefficients in a flapping cycle for a single foil with varying $h_0/c$ (fixed $\theta_0 = 30^{\circ}$): (a) $C_D$ and (b) $C_L$.}
        \label{TH_single_h0_force_CdCl}
\end{figure}

\begin{figure}
		\centering
		\includegraphics[width=\textwidth]{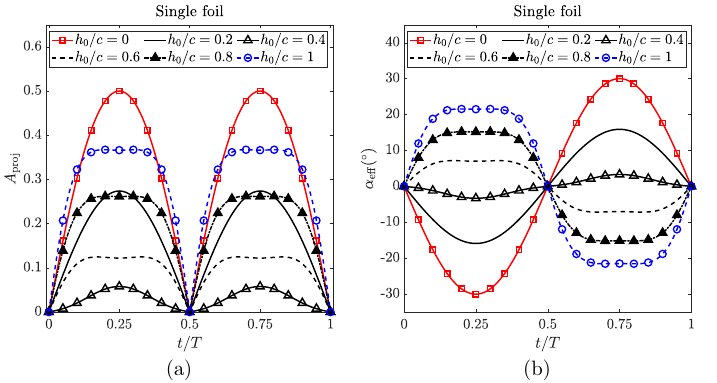}
        \caption{Temporal variation of the following quantities in a flapping cycle for a single foil with varying $h_0/c$ (fixed $\theta_0 = 30^{\circ}$): (a) $A_{\mathrm{proj}}$, and (b) $\alpha_{\mathrm{eff}}$.}
        \label{TH_single_h0_alpha_area}
\end{figure}

The temporal variation of the drag and lift coefficients in a flapping cycle is shown for different heave amplitudes in Fig. \ref{TH_single_h0_force_CdCl}. 
The peak value of the drag coefficient in a flapping cycle first decreases from $h_0/c \in [0, 0.4]$ and then increases with $h_0/c$ wherein the maximum drag coefficient is observed for $h_0/c = 1$. The variation in $C_L$ is quite similar to $C_Y$. However, the amplitude of $C_D$ is smaller compared to $C_L$ depicting lift as the main contributor to the resultant force on the foil. During the downstroke (first-half of the flapping cycle), increasing the heave amplitude increases the heave velocity leading to an increase in the change in the amplitude and direction of the effective velocity $U_\mathrm{eff}$ (see the schematic in Fig. \ref{LEV_single_schematic}(b)). As the direction of the effective velocity changes, so does the projected area of the foil to $U_\mathrm{eff}$. This area has been plotted in Fig. \ref{TH_single_h0_alpha_area}(a).
The variation in the area directly correlates to the drag coefficient as with increasing heave amplitude, the projected area also decreases till $h_0/c = 0.4$ and then increases. On the other hand, the lift coefficient variation is explained by the changing effective angle of attack, depicted in Fig. \ref{TH_single_h0_alpha_area}(b). During the downstroke, the effective angle of attack at a time instant increases with increasing heave amplitude which confirms the increase in $C_L$ at a time instant in the downstroke.

\begin{figure}
		\centering
		\includegraphics[width=\textwidth]{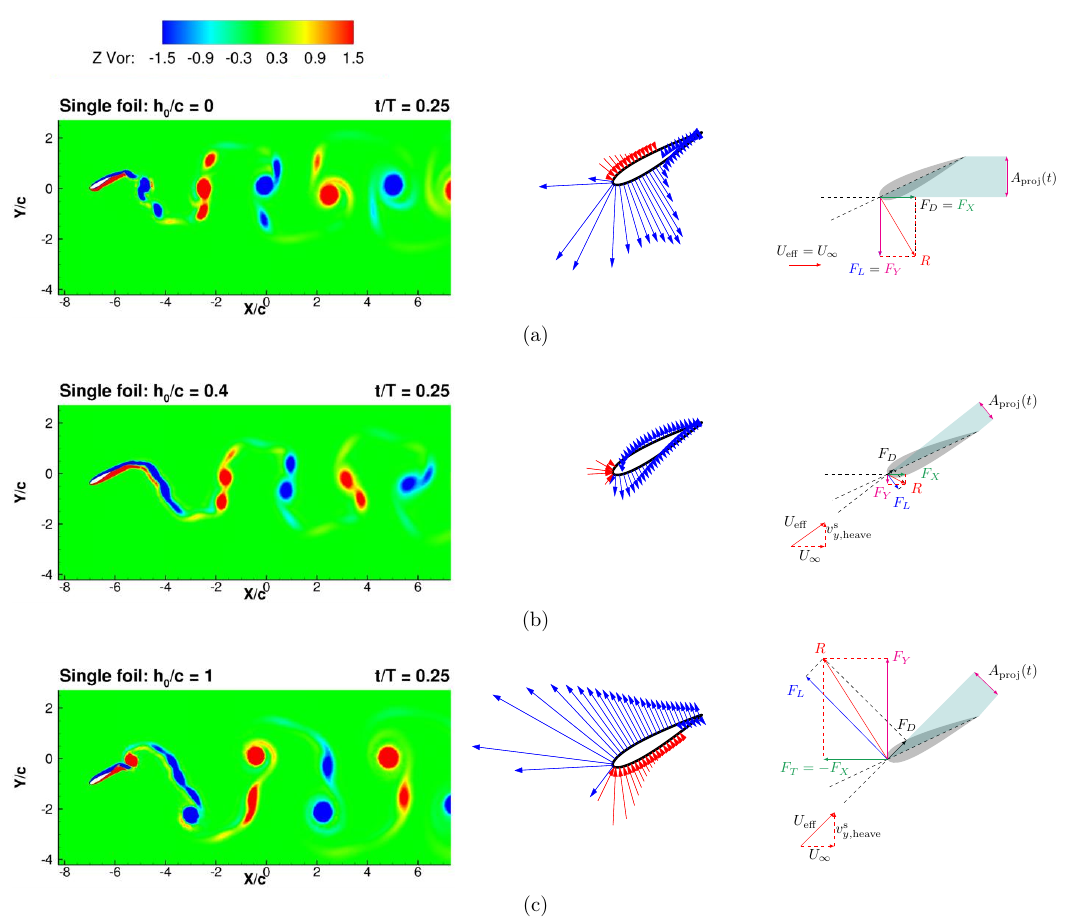}
        \caption{Flapping single foil at $t/T = 0.25$, $\theta_0 = 30^{\circ}$ and (a) $h_0/c = 0$, (b) $h_0/c = 0.4$, and (c) $h_0/c = 1$. The Z-vorticity contours, pressure distribution along the surface of the downstream foil and the instantaneous forces are depicted in the Left, Middle and Right columns, respectively.}
        \label{ZVor_single_h0}
\end{figure}

Next, we look into the wake signature of the flapping foil to confirm our observations regarding the force coefficients and their correlation with the effective angle of attack. The Z-vorticity contour plots (Left) of the wake of the flapping foil along with the pressure distribution (Middle) and forces (Right) are depicted in Fig. \ref{ZVor_single_h0}.
The transition from drag-producing to thrust-producing wake can be noticed in the wake signature of the flapping foil. At $h_0/c = 0$ (Fig. \ref{ZVor_single_h0}(a)), the clockwise (CW) and the counter-clockwise (CCW) vortices (blue and red in color respectively) are aligned such that they produce a von-K$\mathrm{\acute{a}}$rm$\mathrm{\acute{a}}$n (vK) vortex street, which is drag producing. As heave amplitude is increased to $h_0/c = 0.4$ (Fig. \ref{ZVor_single_h0}(b)), the vortex street resembles a scenario close to the transition to thrust-producing inverted von-K$\mathrm{\acute{a}}$rm$\mathrm{\acute{a}}$n (IvK) vortex street (this transition has been referred as ``feathering limit'' in the literature). At $h_0/c = 1$ (Fig. \ref{ZVor_single_h0}(c)), we observe the thrust-producing IvK vortex street. The pressure arrow plots for the same instances of the flapping foil depict the increased suction and positive pressure on the upper and lower surfaces of the foil, respectively with heave amplitude. This results in thrust-favoring conditions as the heave amplitude increases. The forces on the foil at the same instances are also depicted in the figure. With the increase in $h_0/c$, the effective angle of attack increases. This results in the resultant force transitioning such that its X-component gives a more positive thrust with the increase in $h_0/c$. Moreover, the change of the projected area $A_\mathrm{proj}(t)$ can also be noticed in Fig. \ref{ZVor_single_h0}(Right) which decreases from $h_0/c = 0$ to $h_0/c = 0.4$ and then increases for $h_0/c = 1$. As mentioned earlier, this translates directly to the trend in the drag coefficient in Fig. \ref{TH_single_h0_force_CdCl}(a). 

The main contributor to the resultant force in this case is the lift force. Thus, for the parameters considered, an increase in the thrust coefficient with heave amplitude is noticed as a consequence of increase in the effective angle of attack.
% (effective area does not vary with change in heave amplitude).
This dependence of the thrust coefficient on the effective angle of attack for heaving foil corroborates the observations by \citet{vanBuren2019}, where it was pointed out that the thrust is entirely a consequence of lift forces. At fixed pitch amplitude and $f^*$, the mean thrust coefficient was found to increase with heave amplitude. The study conducted by \citet{Yu2017} observed a similar effect of changing the heave amplitude for fixed flapping frequency and pitch amplitude.

\subsection{Effect of pitch amplitude $(\theta_0)$ on propulsion}
\label{single_theta0_effects}
\begin{figure}
		\centering
		\includegraphics[width=\textwidth]{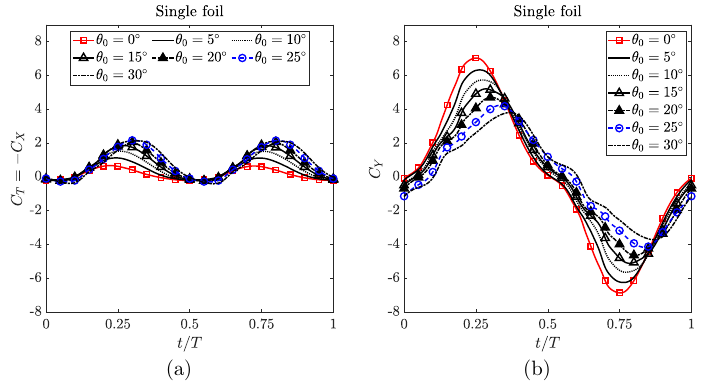}
        \caption{Temporal variation of the force coefficients in a flapping cycle for a single foil with varying $\theta_0$ (fixed $h_0/c = 1$): (a) $C_T$, and (b) $C_Y$.}
        \label{TH_single_theta0_force_CtCy}
\end{figure}

For observing the variation of the propulsive performance with the pitch amplitude, the heave amplitude is fixed at $h_0/c = 1$ along with the Reynolds number $Re=1100$ and the flapping frequency $f^* = 0.2$. The pitch amplitude is varied between $0^{\circ}$ and $30^{\circ}$ with increments of $5^{\circ}$. The time history of the thrust coefficient is shown in Fig. \ref{TH_single_theta0_force_CtCy}(a). It is seen that the maximum thrust coefficient in a flapping cycle increases with an increase in the pitch amplitude, similar to the observations by \citet{Yu2017} for low frequencies. Furthermore, the instant of the peak thrust coefficient is delayed as the pitch amplitude increases. 

\begin{figure}
		\centering
		\includegraphics[width=\textwidth]{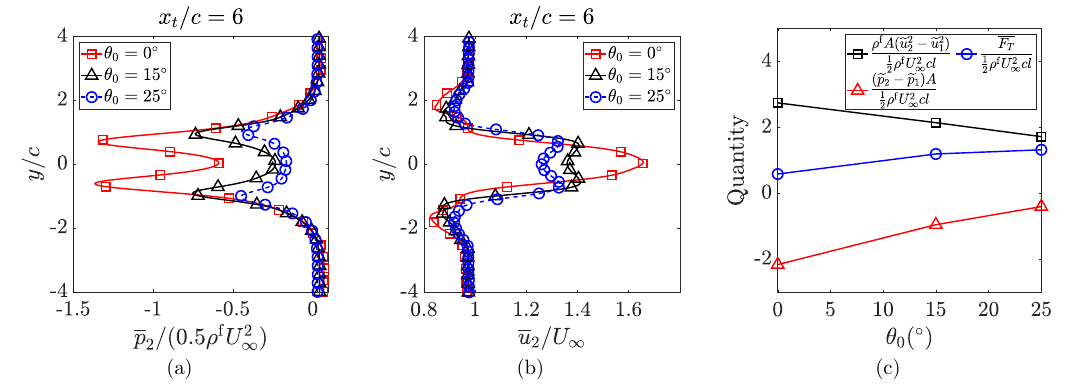}
        \caption{Control volume analysis of the flapping foil for the effect of pitch amplitude: (a) time-averaged pressure at the right boundary $(\overline{p}_2)$, (b) time-averaged streamwise velocity at the right boundary $(\overline{u}_2)$, and (c) evaluated quantities based on Eq. (\ref{Mom_Bal_Eq}) for varying $\theta_0$.}
        \label{Single_theta0_CV_quantity}
\end{figure}

It can also be deduced that the mean thrust coefficient is positive for all $\theta_0$ values, however it increases with $\theta_0$. A control volume analysis similar to the previous subsection (Fig. \ref{Single_CV}) can be performed here, with variations in $\overline{p}_2$ and $\overline{u}_2$ at the right boundary of the control volume shown in Fig. \ref{Single_theta0_CV_quantity}. Here, we observe a peculiar behavior. Although the velocity $\widetilde{u}_2$ is higher for $\theta_0 = 0^{\circ}$, it has the lowest mean thrust among the cases considered. Reviewing the conservation of linear momentum derived in Eq. (\ref{Mom_Bal_Eq}), it can be observed that in this scenario, the second term $(\widetilde{p}_2 - \widetilde{p}_1)A$ has significant effect as higher negative pressures are observed for $\theta_0 = 0^{\circ}$ case, in comparison to others (depicted in Fig. \ref{Single_theta0_CV_quantity}(c)). Therefore, it is important to consider the pressure distribution at the wake of the foil to get a clear understanding of the thrust generation. The time-averaged streamwise velocity pattern solely does not paint a clear picture of the average thrust. As can be observed from Fig. \ref{Single_theta0_CV_quantity}(c), there are competing effects of the two terms, viz., $(\widetilde{p}_2 - \widetilde{p}_1)A$ and $\rho^\mathrm{f} A (\widetilde{u}_2^2 - \widetilde{u}_1^2)$, which result in the trend for the generated thrust.
%\begin{figure}
%		\centering
%		\includegraphics[width=\textwidth]{control_volume_theta0.pdf}
%        \caption{Control volume analysis of the flapping foil for the effect of pitch amplitude: The inset plots indicate the time-averaged pressure and streamwise velocities at the left and right ends of the control volume.}
%        \label{Single_theta0_CV}
%\end{figure}

\begin{figure}
		\centering
		\includegraphics[width=\textwidth]{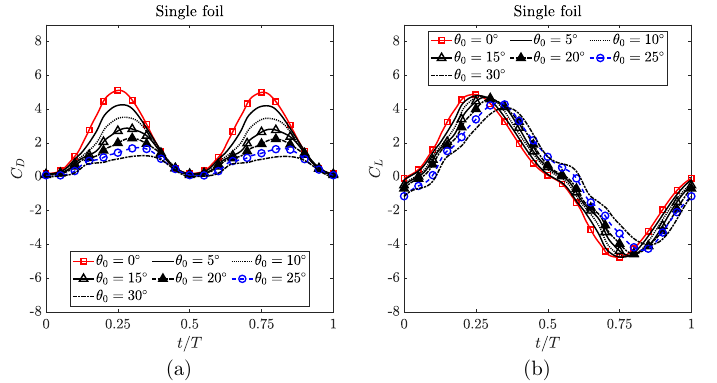}
        \caption{Temporal variation of the force coefficients in a flapping cycle for a single foil with varying $\theta_0$ (fixed $h_0/c = 1$): (a) $C_D$, and (b) $C_L$.}
        \label{TH_single_theta0_force_CdCl}
\end{figure}

\begin{figure}
		\centering
		\includegraphics[width=\textwidth]{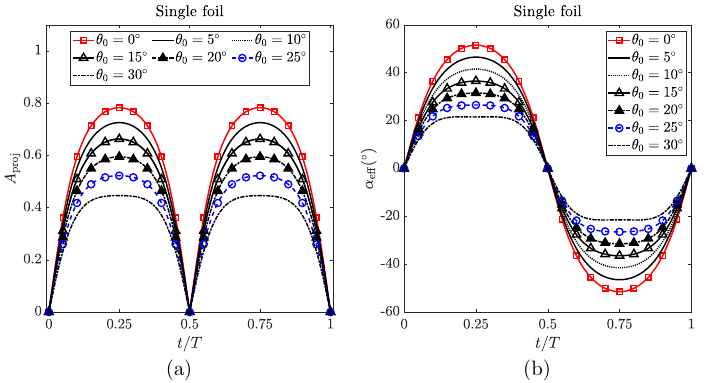}
        \caption{Temporal variation of the following quantities in a flapping cycle for a single foil with varying $\theta_0$ (fixed $h_0/c = 1$): (a) $A_{\mathrm{proj}}$, and (b) $\alpha_{\mathrm{eff}}$.}
        \label{TH_single_theta0_alpha_area}
\end{figure}

The interplay of the drag and lift forces can be observed in Fig. \ref{TH_single_theta0_force_CdCl}. 
In contrast to the previous subsection, here the drag force is noted to have significant contribution to the resultant force, especially for smaller $\theta_0$ values. The peak value of the drag coefficient decreases with increasing pitch amplitude. As the heave amplitude is constant for the cases considered, the effective velocity $U_\mathrm{eff}$ remains at the same amplitude and direction across the various pitch amplitudes. However, as the pitch amplitude changes, so does the projected area of the foil to the flow direction, as shown in Fig. \ref{TH_single_theta0_alpha_area}(a).
The variation in the projected area translates to the drag coefficient plot. The lift coefficient (Fig. \ref{TH_single_theta0_force_CdCl}(b)) has a similar variation to $C_Y$. Minor changes are observed across the different pitch amplitudes, with highest lift noted for $\theta_0 = 0^{\circ}$ during the downstroke. The lift coefficient can be correlated with the changing effective angle of attack across pitch amplitudes, depicted in Fig. \ref{TH_single_theta0_alpha_area}(b).
 
%To comprehend this relationship, the effective angle of attack and the effective area are depicted in Figs. \ref{TH_single_theta0}(b) and \ref{TH_single_theta0}(c), respectively. Contrary to the observation in the previous sub-section, the effective angle of attack $\alpha_{\mathrm{eff}}$ during the downstroke is positive but the peak of $\alpha_{\mathrm{eff}}$ decreases with increase in the pitch amplitude. This trend of $\alpha_{\mathrm{eff}}$ does not translate to the thrust coefficient plot. This can be attributed to the increase in the effective area projected to the incoming flow with increase in the pitch amplitude (Fig. \ref{TH_single_theta0}(c)). The increased projected area leads to an increase in net thrust as a consequence of suction pressure on the upper surface of the foil and high pressure on the lower surface of the foil during downstroke. Therefore, in this scenario, the effective projected area is the prominent factor driving the propulsive performance.

\begin{figure}
		\centering
		\includegraphics[width=\textwidth]{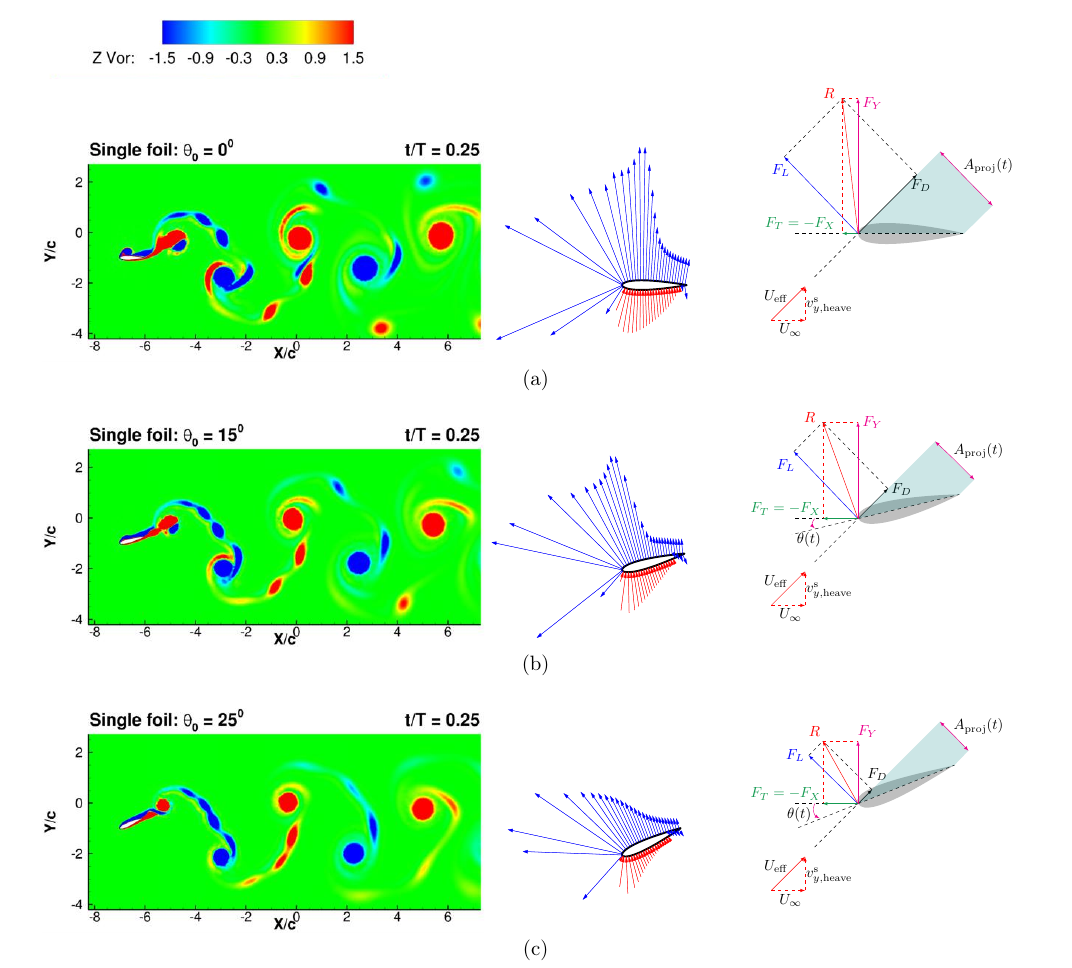}
        \caption{Flapping single foil at $t/T = 0.25$, $h_0/c = 1$ and (a) $\theta_0 = 0^{\circ}$, (b) $\theta_0 = 15^{\circ}$, and (c) $\theta_0 = 25^{\circ}$. The Z-vorticity contours, pressure distribution along the surface of the downstream foil and the instantaneous forces are depicted in the Left, Middle and Right columns, respectively.}
        \label{ZVor_single_theta0}
\end{figure}

The wake signature of the flapping foil is visualized by Z-vorticity contours in Fig. \ref{ZVor_single_theta0} along with pressure distribution and the forces on the foil. As all the pitch amplitudes produce net thrust, an IvK vortex street is observed for all the cases. Furthermore, with increase in pitch amplitude, the width of the wake increases which can be confirmed by the time-averaged streamwise velocity plot in Fig. \ref{Single_theta0_CV_quantity}(b). The pressure distribution across the various pitch amplitudes more or less remain the same. However, it is the inclination of the foil with the horizontal which changes, leading to a decrease in the projected area of the foil to the incoming flow. The inclination due to the increasing pitch amplitude also inclines the resultant force towards the negative X direction, increasing the thrust component of the force. Another perspective could be the increase in the frontal area of the foil to the freestream velocity $U_{\infty}$ through which the pressure differential across the upper and lower surfaces of the foil tends to give more thrust component as pitch amplitude increases.
%Therefore, for the parameters considered, an increase in the thrust coefficient with pitch amplitude is observed as a consequence of increase in the effective projected area and increased width of the wake. In addition, the observation of the vortex pattern suggests that the LEV is more prominent for lower $\theta_0$ as the effective angle of attack is higher.

The key takeaways from the analysis of the single flapping foil are:

$\bullet$  Effective angle of attack $(\alpha_\mathrm{eff}(t))$ relates to the lift force and the projected area to the incoming flow $(A_\mathrm{proj}(t))$ relates to the drag force.

$\bullet$ Average thrust increases with the heave amplitude with majority contribution from the lift force which increases with increasing effective angle of attack.

$\bullet$ The mean thrust increases with the pitch amplitude, while decreasing the projected area of the foil to the incoming flow $U_\mathrm{eff}$. On the other hand, the projected area to the freestream velocity $U_{\infty}$ increases resulting in higher component of thrust force.

$\bullet$ The behavior of both the time-averaged pressure and streamwise velocity at the wake of the foil are necessary to quantify the mean thrust. 

With this background context about the influence of the kinematic parameters on the propulsive characteristics of a single isolated flapping foil, we next consider the more complex tandem configuration.

%================================================================================
\section{Flapping foils in tandem configuration}
\label{tandem_flapping_foil}

In the tandem configuration of flapping foils, the downstream foil interacts with the wake of the upstream foil resulting in a very complex interaction. Unlike the single isolated foil discussed in the previous section, the flow physics is not that straightforward to understand in this scenario. It is very difficult to define the basic parameters such as effective angle of attack and the projected area for the downstream foil, as the direction of incoming flow is disturbed by the upstream foil. Therefore, we focus on the vortex interaction mechanisms to comprehend the behavior of the downstream foil under the influence of the upstream foil's wake. We also consider the projected frontal area of the flapping foil to the freestream velocity $U_{\infty}$ in our discussion. The effect of the gap between the flapping foils on the propulsive performance of the downstream foil was investigated in detail by \citet{Joshi2021} for gap ratios of $g/c_d \in [1, 14]$ for foils of identical chord length, i.e., $c_u/c_d = 1$. Various mechanisms of vortex interactions were proposed and categorized into constructive and destructive interactions, which have been noticed by \citet{Broering2012, gopalkrishnan_triantafyllou_triantafyllou_barrett_1994, Lewin_2003, muscutt_weymouth_ganapathisubramani_2017}.

\begin{figure}
	\centering
    \includegraphics{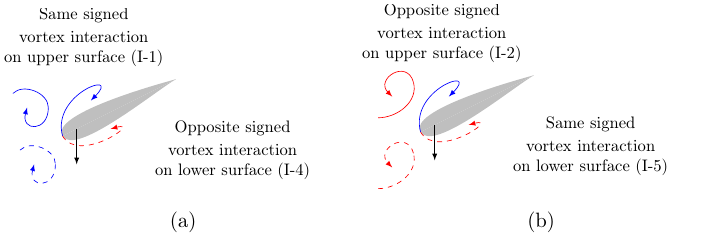}
\caption{Major types of interactions of the downstream foil with the upstream foil's wake for (a) favorable, and (b) unfavorable, thrust generating conditions in the work by \citet{Joshi2021}.}
	\label{LEV_tandem_schematic}
\end{figure}

A constructive interaction occurs when same-signed vortices interact, leading to supply of vorticity on the surface of the downstream foil. On the other hand, a destructive interaction involves the interaction of opposite-signed vortices which drives away the vorticity on the surface of the foil. It is the combination of above interactions on the upper and lower surfaces of the foil which determines the favorable and unfavorable conditions for generation of thrust. Major interactions have been shown in Fig. \ref{LEV_tandem_schematic}, the details of which can be found in the work by \citet{Joshi2021}. To summarize, a constructive interaction (same-signed) on the upper surface (Interaction-1 or I-1) and a destructive one (opposite-signed) on the lower surface (Interaction-4 or I-4) of the foil favors the generation of thrust during the downstroke. On the contrary, a destructive interaction on the upper surface (Interaction-2 or I-2) and a constructive one on the lower surface (Interaction-4 or I-4) leads to unfavorable thrust generating conditions. Furthermore, when an opposite-signed vortex travels in proximity to the upper surface of the foil, it tends to pull the shear layer leading to an enlargement of LEV, which is favorable (Interaction-3 or I-3) (not shown in Fig. \ref{LEV_tandem_schematic}). The earlier (later) the favorable (unfavorable) condition occurs during the downstroke, the better is the thrust generation for the downstream foil in the tandem arrangement.
% Fig 9

Based on the gap ratios and the considered parameters in \citet{Joshi2021} ($Re=1100$, $c_u/c_d = 1$, $\theta_0^u = \theta_0^d = 30^{\circ}$, $f^*_u = f^*_d = 0.2$,  $h_0^u/c_d = h_0^d/c_d = 1$, $\phi^u_h = \phi^d_h = 90^{\circ}$), a periodic variation in the thrust and propulsive efficiency was noticed with the increase in gap ratio, similar to the observations in the literature. The gap ratio of $g/c_d = 4$ manifested lowest mean thrust coefficient (most unfavorable condition) and $g/c_d = 7$ depicted the highest thrust coefficient (most favorable condition). Therefore, in the current investigation to study the effects of heave and pitch amplitudes on the propulsive performance, we consider these two representative gap ratios. We perform two-dimensional computations of the tandem flapping foils to investigate the effects of kinematic motion on propulsion. 

\subsection{Effect of heave amplitude of upstream foil $(h_0^u)$ on propulsion}
\begin{figure}
		\centering
		\includegraphics[width=\textwidth]{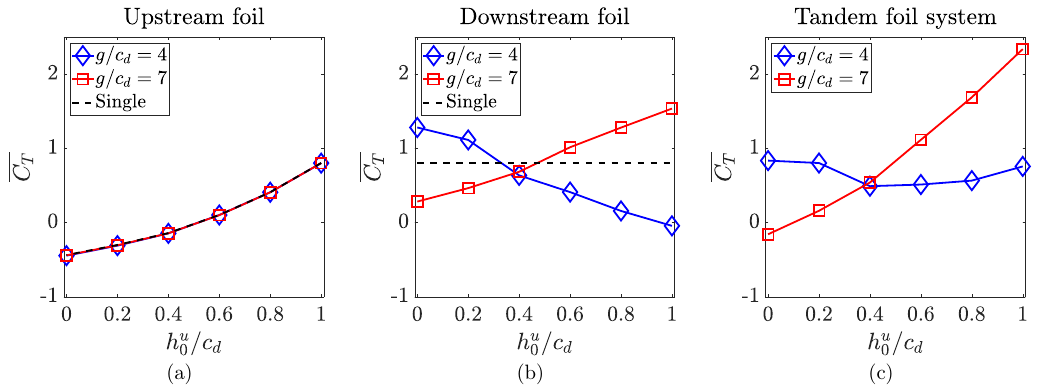}
        \caption{Variation of the mean thrust coefficient for the tandem flapping foils at $h^d_0/c_d = 1$, $\theta_0^u = \theta_0^d = 30^{\circ}$ with varying upstream foil heave amplitude $h^u_0/c_d$ for (a) upstream foil, (b) downstream foil, and (c) tandem foil system.}
        \label{CTmean_h0u}
\end{figure}

\begin{figure}
		\centering
		\includegraphics[width=\textwidth]{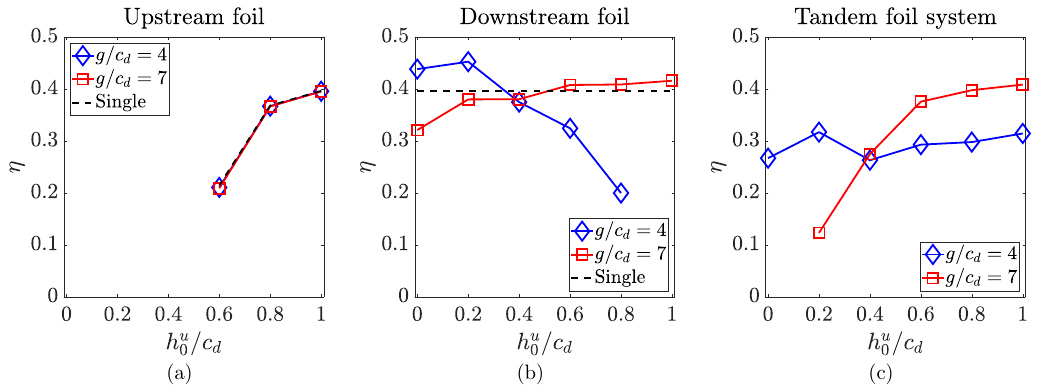}
        \caption{Variation of the propulsive efficiency $\eta$ for the tandem flapping foils at $h^d_0/c_d = 1$, $\theta_0^u = \theta_0^d = 30^{\circ}$ with varying upstream foil heave amplitude $h^u_0/c_d$ for (a) upstream foil, (b) downstream foil, and (c) tandem foil system.}
        \label{Eta_h0u}
\end{figure}

Here, we discuss the influence of the heave amplitude of the upstream foil on the thrust and propulsive efficiency of the tandem foils. As mentioned earlier, two gap ratios of $g/c_d = 4, 7$ are considered. The downstream heave amplitude is fixed ($h_0^d/c_d = 1$) and both the upstream and downstream pitch amplitudes are held constant ($\theta_0^u = \theta_0^d = 30^\circ$). As expected, the mean thrust coefficient of the upstream foil follows the trend of the single foil, indicating null interference of the downstream foil on the upstream foil characteristics, as shown in Fig. \ref{CTmean_h0u}(a). An increase in $\overline{C_T}$ is observed with $h_0^u/c_d$ which has been discussed in detail in the previous section pertaining to a single flapping foil. For the downstream foil, the variation in mean thrust is shown in Fig. \ref{CTmean_h0u}(b). The mean thrust decreases and increases with $h_0^u/c_d$ for $g/c_d = 4$ and $g/c_d = 7$, respectively. For the combined tandem system, the mean thrust coefficient is depicted in Fig. \ref{CTmean_h0u}(c). For the gap ratio of $7$, the combined $\overline{C_T}$ is observed to increase with $h_0^u/c_d$, reaching a maximum value of 2.34 at $h_0^u/c_d = 1$. In contrast, at $g/c_d = 4$, the thrust decreases slightly at $h_0^u/c_d = 0.4$ and then increases marginally. The maximum mean thrust for $g/c_d = 4$ is 0.84 at $h_0^u/c_d = 0$. The propulsive efficiency is depicted in Fig. \ref{Eta_h0u}. The efficiency for the upstream foil is identical to the isolated single foil. It increases sharply as the dimensionless heave amplitude grows from 0.6 to 0.8 and continues to increase at higher amplitudes albeit slowly. For $h_0^u/c_d = [0, 0.2, 0.4]$, negative thrust or drag is observed for the upstream foil and therefore, the efficiency has not been plotted. For the downstream foil, a decrease in efficiency with increasing $h_0^u/c_d$ is observed for $g/c_d = 4$, while it increases gradually for $g/c_d = 7$. Note that in Figs. \ref{CTmean_h0u}(b) and \ref{Eta_h0u}(b), the ``single'' foil depicts the results for an isolated foil with $h_0/c = 1$ as the heave amplitude of the downstream foil is also unchanged, i.e., $h_0^d/c_d = 1$. For the combined tandem foils shown in Fig. \ref{Eta_h0u}(c), the propulsive efficiency follows a similar trend as that of the mean thrust coefficient. The maximum efficiency of 41\% is observed for $g/c_d = 7$ at $h_0^u/c_d = 1$ and about 32\% for $g/c_d = 4$ at $h_0^u/c_d = 0.2$.

% Fig 10
% Fig 11

The average thrust coefficient variation with $h_0^u/c_d$ can be realized by revisiting the control volume analysis discussed in Section \ref{single_h0_effects} for a single foil. Here, we will apply the conservation of linear momentum in the streamwise direction on a control volume enclosing the tandem foil configuration. The mean thrust force is given by $\overline{F_T} = \rho^\mathrm{f}A(\widetilde{u}_2^2 - \widetilde{u}_1^2) + (\widetilde{p}_2 - \widetilde{p}_1)A$. We have observed that the variation in the freestream quantities such as $\widetilde{p}_1$ and $\widetilde{u}_1$ on the left boundary of the control volume for various kinematic parameters is negligible. On the other hand, the quantities $\widetilde{p}_2$ and $\widetilde{u}_2$ on the right boundary of the control volume depict the averaged wake of the tandem system and are of importance to the study. These quantities are evaluated by integrating the time-averaged pressure $(\overline{p}_2)$ and velocity $(\overline{u}_2)$ across the Y-direction similar to Eqs. (\ref{int_p1}) and (\ref{int_u1}).

\begin{figure}
		\centering
		\includegraphics[width=\textwidth]{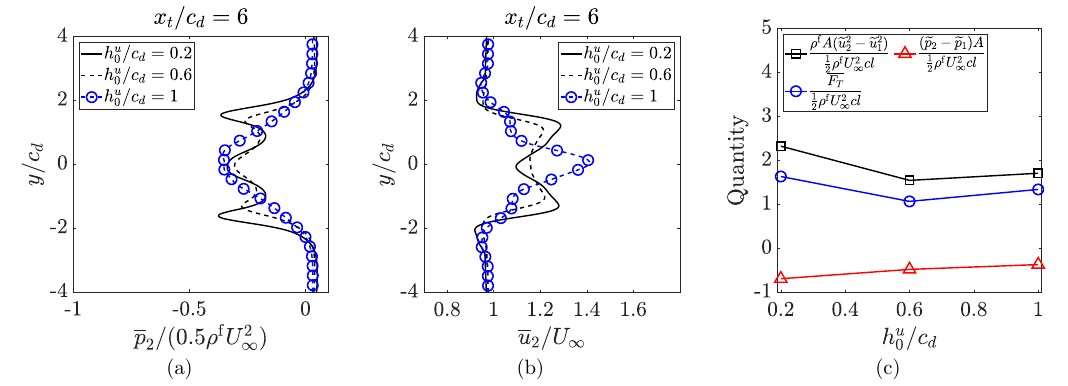}
        \caption{Control volume analysis of the tandem system of foils: (a) time-averaged pressure $\overline{p}_2$, (b) time-averaged streamwise velocity $\overline{u}_2$, at the right boundary of the control volume; and (c) evaluated quantities based on Eq. (\ref{Mom_Bal_Eq}) for representative $h_0^u/c_d$; for the gap ratio of $g/c_d = 4$.}
        \label{g4_pAvg_UAvg_h0u}
\end{figure}

\begin{figure}
		\centering
		\includegraphics[width=\textwidth]{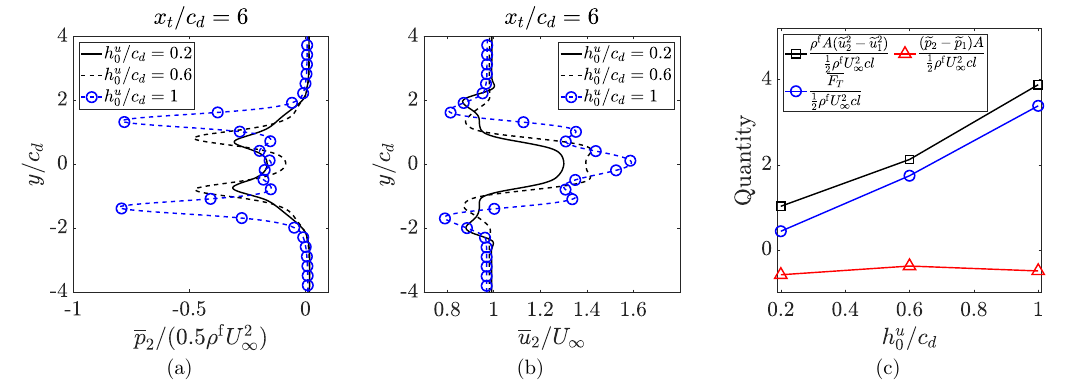}
        \caption{Control volume analysis of the tandem system of foils: (a) time-averaged pressure $\overline{p}_2$, (b) time-averaged streamwise velocity $\overline{u}_2$, at the right boundary of the control volume; and (c) evaluated quantities based on Eq. (\ref{Mom_Bal_Eq}) for representative $h_0^u/c_d$; for the gap ratio of $g/c_d = 7$.}
        \label{g7_pAvg_UAvg_h0u}
\end{figure}

The variation in $\overline{p}_2$ and $\overline{u}_2$ at a distance of $x_t = 6c_d$ from the leading edge of the downstream foil is shown in Figs. \ref{g4_pAvg_UAvg_h0u}(a-b) and \ref{g7_pAvg_UAvg_h0u}(a-b) for gap ratio of 4 and 7, respectively. The average pressure distribution across the right boundary of the control volume is such that the quantity $(\widetilde{p}_2 - \widetilde{p}_1)A$ is almost identical across the different heave amplitudes, which is observed in Figs. \ref{g4_pAvg_UAvg_h0u}(c) and \ref{g7_pAvg_UAvg_h0u}(c). Therefore, the average streamwise velocity distribution $\overline{u}_2$ dictates the mean thrust of the tandem system in this scenario which is clear from the evaluated quantity $\rho^\mathrm{f}A(\widetilde{u}_2^2 - \widetilde{u}_1^2)$ and the trend in $\overline{F_T}$ in the figure.

\begin{figure}
		\centering
		\includegraphics[width=\textwidth]{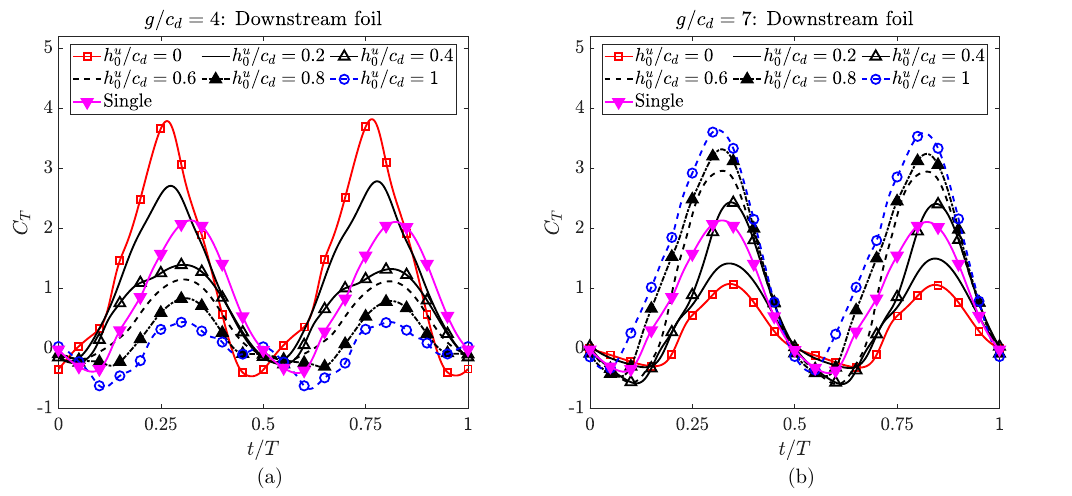}
        \caption{Temporal variation of the thrust coefficient for the downstream foil for varying $h_0^u/c_d$ at $g/c_d$: (a) 4, and (b) 7.}
        \label{TH_CT_h0u}
\end{figure}

\begin{figure}
		\centering
		\includegraphics[width=\textwidth]{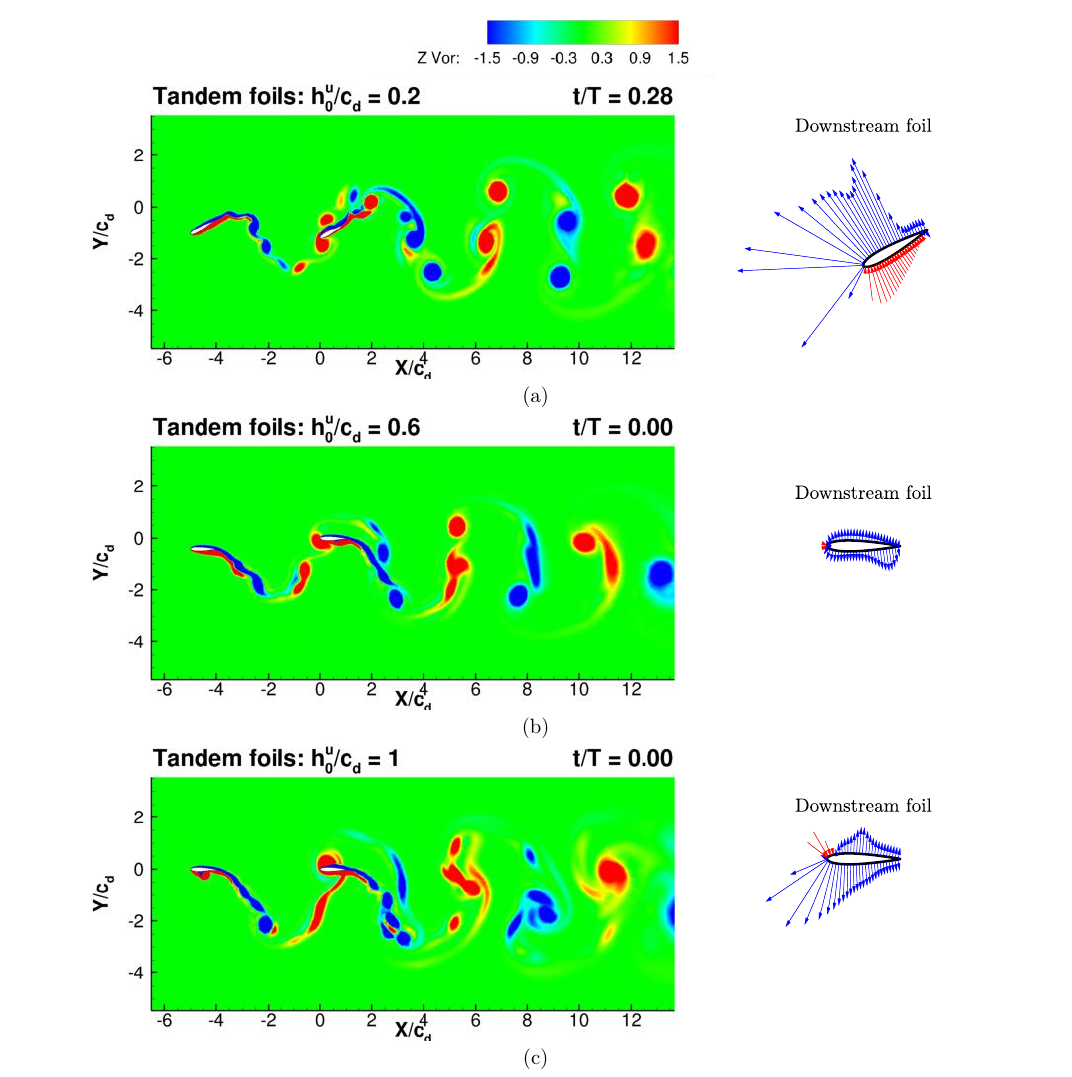}
        \caption{Flapping foils at different time instances of a flapping cycle at $g/c_d = 4$, $h^d_0/c_d = 1$, $\theta_0^u = \theta_0^d = 30^{\circ}$ and $h_0^u/c_d$: (a) 0.2, (b) 0.6, and (c) 1. The Z-vorticity contours and pressure distribution along the surface of the downstream foil are depicted on the Left and Right column at the instant.}
        \label{ZVor_tandem_g4_h0u}
\end{figure}

\begin{figure}
		\centering
		\includegraphics[width=\textwidth]{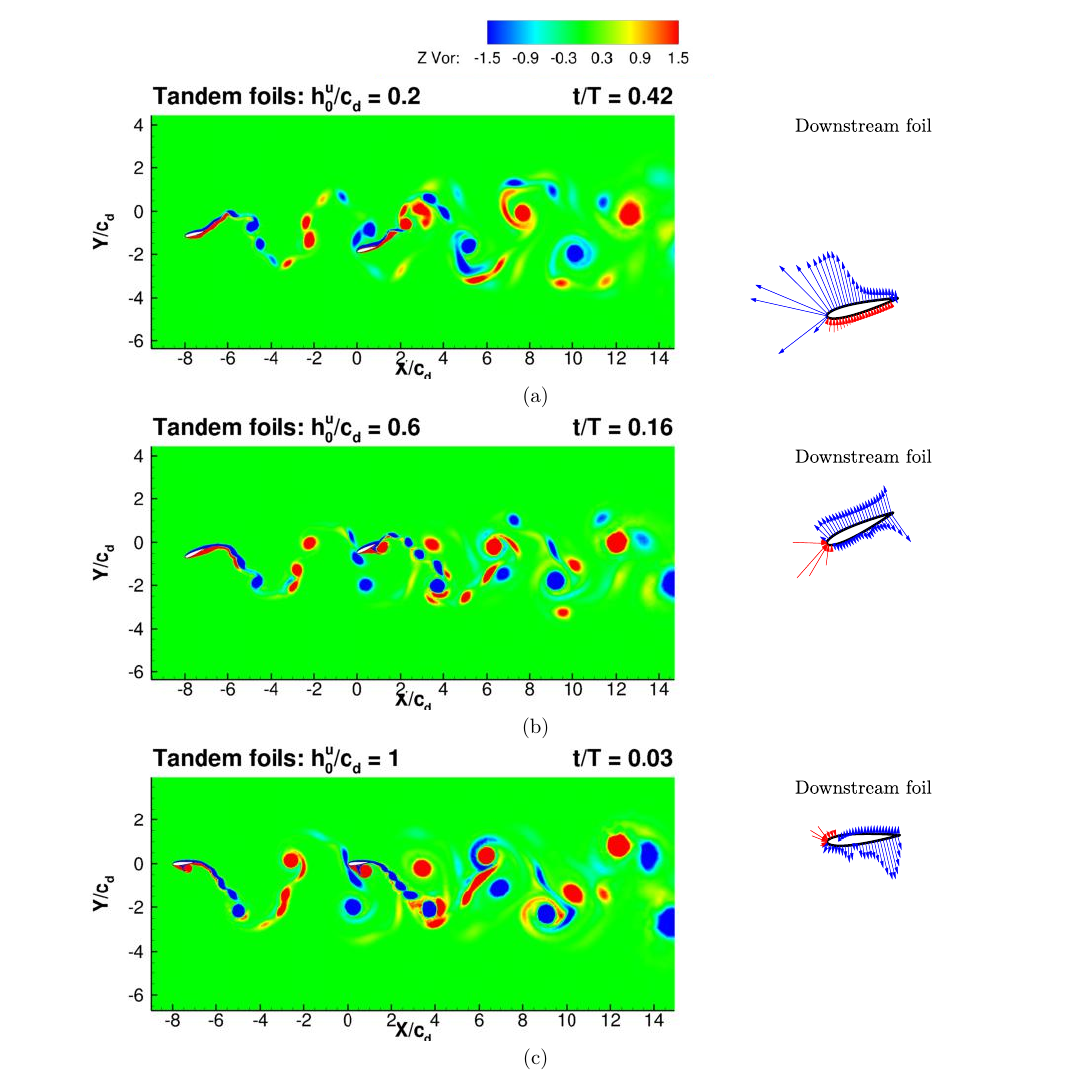}
        \caption{Flapping foils at different time instances of a flapping cycle at $g/c_d = 7$, $h^d_0/c_d = 1$, $\theta_0^u = \theta_0^d = 30^{\circ}$ and $h_0^u/c_d$: (a) 0.2, (b) 0.6, and (c) 1. The Z-vorticity contours and pressure distribution along the surface of the downstream foil are depicted on the Left and Right column at the instant.}
        \label{ZVor_tandem_g7_h0u}
\end{figure}

The variation of the thrust coefficient of the downstream foil for the two gap ratios under various $h_0^u/c_d$ is shown in Fig. \ref{TH_CT_h0u}. It can be observed that the thrust generation capability of the downstream foil decreases with increasing $h_0^u/c_d$ for $g/c_d = 4$, while a contrasting effect is observed for high gap ratio of 7. These transient variations in thrust are a result of favorable or unfavorable interactions of the wake of the upstream foil with the downstream foil.
%The effect of the tandem configuration on the propulsive performance of the downstream foil can be investigated further by observing the vortex interaction mechanism of the upstream foil's wake with the downstream foil. 
This is shown in the wake signature diagram by visualizing the Z-vorticity contours at different instances for the gap ratios $g/c_d = 4$ and $g/c_d = 7$ in Figs. \ref{ZVor_tandem_g4_h0u} and \ref{ZVor_tandem_g7_h0u}, respectively. For $g/c_d = 4$, at $h_0^u/c_d = 0.2$, the CCW vortex just passes the upper surface of the foil in close proximity at $t/T = 0.28$ (I-3) which leads to the LEV getting pulled on the upper surface during the downstroke (see Fig. \ref{ZVor_tandem_g4_h0u}(a)). This gives the favorable condition for thrust generation near the mid-stroke. The negative and positive pressure on the upper and lower surface of the downstream foil respectively is represented by the arrow diagram in the figure. As the heave amplitude is increased to $h_0^u/c_d = 0.6 - 1$, the interaction becomes unfavorable as the CCW vortex interacts with the upper and lower surface of the foil fairly early in the downstroke (see Figs. \ref{ZVor_tandem_g4_h0u}(b) and (c)). This early interaction takes away the CW vorticity from the upper surface eventually leading to premature shedding of the LEV. Furthermore, the interaction also supplies the CCW vorticity to the lower surface of the foil strengthening the shear layer resulting in suction pressure on the lower surface (Figs. \ref{ZVor_tandem_g4_h0u}(b) and (c)).
% Fig 12

At $g/c_d = 7$, the variation of the mean thrust coefficient is opposite to that of $g/c_d = 4$. Although all the $h_0^u/c_d$ cases generate positive thrust or are favorable conditions; with increase in $h_0^u/c_d$, the interaction of the CW vorticity with the downstream foil occurs earlier (I-1 and I-4) during the downstroke, as can be seen in Fig. \ref{ZVor_tandem_g7_h0u}. For example, this interaction occurs at $t/T = 0.42$ (almost the end of downstroke) for $h_0^u/c_d = 0.2$ (Fig. \ref{ZVor_tandem_g7_h0u}(a)) but occurs at the start of the downstroke $t/T = 0.03$ for $h_0^u/c_d = 1$ (Fig. \ref{ZVor_tandem_g7_h0u}(c)). This earlier interaction generates a larger LEV (as vorticity is supplied due to constructive interaction on the upper surface) during the prime configuration of the downstroke where the projected area of the foil to the freestream direction is the largest, thus leading to higher thrust. 
% Fig 13

\subsection{Effect of heave amplitude of downstream foil $(h_0^d)$ on propulsion}
Next, we understand the effect of varying the heave amplitude of the downstream foil on its propulsive performance. To accomplish this, we fix the heave amplitude of the upstream foil $h^u_0/c_d = 1$, the pitch amplitudes as $\theta_0^u = \theta_0^d = 30^{\circ}$ and the non-dimensional flapping frequency is selected as $f^* = 0.2$. The heave amplitude of the downstream foil is varied in the range $h_0^d/c_d \in [0, 1]$.

\begin{figure}
		\centering
		\includegraphics{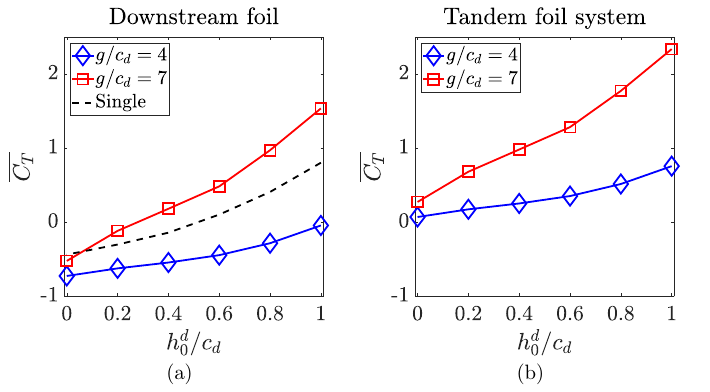}
        \caption{Variation of the mean thrust coefficient for the (a) downstream foil, and (b) tandem foil system, at $h^u_0/c_d = 1$, $\theta_0^u = \theta_0^d = 30^{\circ}$ with varying downstream foil heave amplitude $h^d_0/c_d$.}
        \label{CTmean_h0d}
\end{figure}

\begin{figure}
		\centering
		\includegraphics{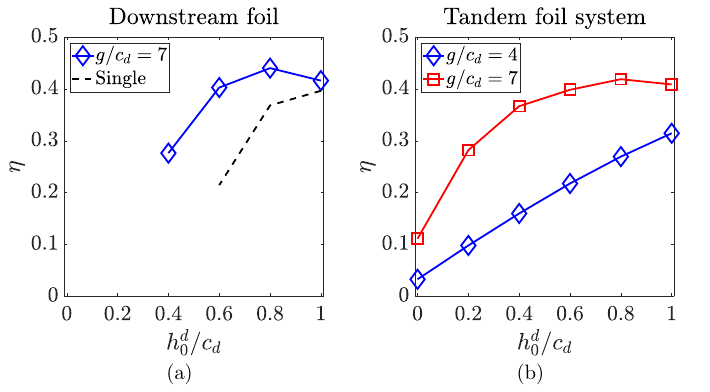}
        \caption{Variation of the propulsive efficiency for the (a) downstream foil, and (b) tandem foil system, at $h^u_0/c_d = 1$, $\theta_0^u = \theta_0^d = 30^{\circ}$ with varying downstream foil heave amplitude $h^d_0/c_d$.}
        \label{Eta_h0d}
\end{figure}

The variation in the mean thrust coefficient and the propulsive efficiency for the tandem foils are shown in Figs. \ref{CTmean_h0d} and \ref{Eta_h0d}, respectively. Note that as the upstream heave amplitude is fixed, the propulsive performance of the upstream foil will be identical to an isolated single foil at $h_0/c = 1$, $\theta_0 = 30^{\circ}$ and $f^* = 0.2$. It is observed that the mean thrust for the downstream foil decreases with decrease in the heave amplitude of the downstream foil for both the gap ratios, as shown in Fig. \ref{CTmean_h0d}(a), similar to the variation for a single foil with decreasing $h_0/c$. At $g/c_d = 4$, we do not observe a positive thrust coefficient for the downstream foil for the heave amplitudes considered. On the other hand, for $g/c_d = 7$, the thrust coefficient is negative for $h_0^d/c_d = [0, 0.2]$. For the combined system (Fig. \ref{CTmean_h0d}(b)), the mean thrust increases for both the gap ratios with increasing $h_0^d/c_d$, similar to the behavior of the downstream foil as the thrust of the upstream foil (constant) gets added to that of the downstream foil in the combined system. The propulsive efficiency for the downstream foil, shown in Fig. \ref{Eta_h0d}(a), first increases slightly when $h_0^d/c_d$ is decreased from 1 to 0.8, and then decreases monotonically for $g/c_d = 7$. As the average thrust of the combined system is always positive, the propulsive efficiency is defined for all the cases, as depicted in Fig. \ref{Eta_h0d}(b). For $g/c_d = 4$, the efficiency is observed to increase linearly with $h_0^d/c_d$, reaching a maximum value of 31.4\% at $h_0^d/c_d = 1$. The behavior of the efficiency is nonlinear in nature and has a maximum value of 42\% at $h_0^d/c_d = 0.8$ for $g/c_d = 7$.
% Fig 14
% Fig 15

\begin{figure}
		\centering
		\includegraphics[width=\textwidth]{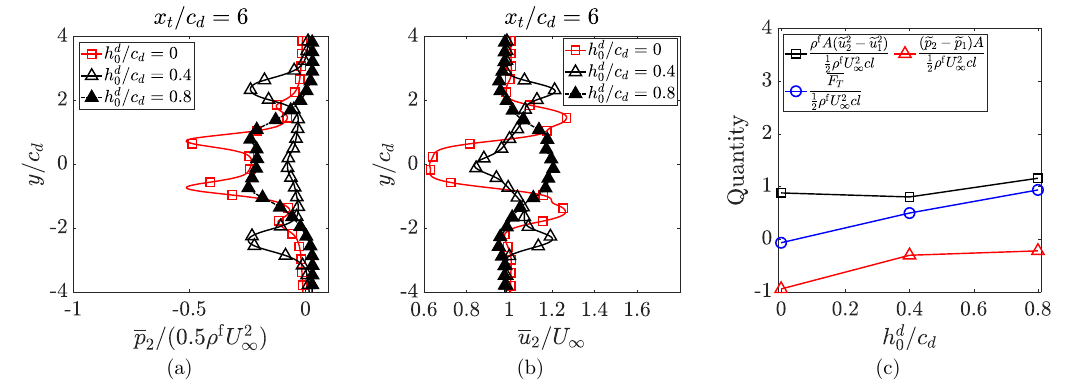}
        \caption{Control volume analysis of the tandem system of foils: (a) time-averaged pressure $\overline{p}_2$, (b) time-averaged streamwise velocity $\overline{u}_2$, at the right boundary of the control volume; and (c) evaluated quantities based on Eq. (\ref{Mom_Bal_Eq}) for representative $h_0^d/c_d$; for the gap ratio of $g/c_d = 4$.}
        \label{g4_pAvg_UAvg_h0d}
\end{figure}

\begin{figure}
		\centering
		\includegraphics[width=\textwidth]{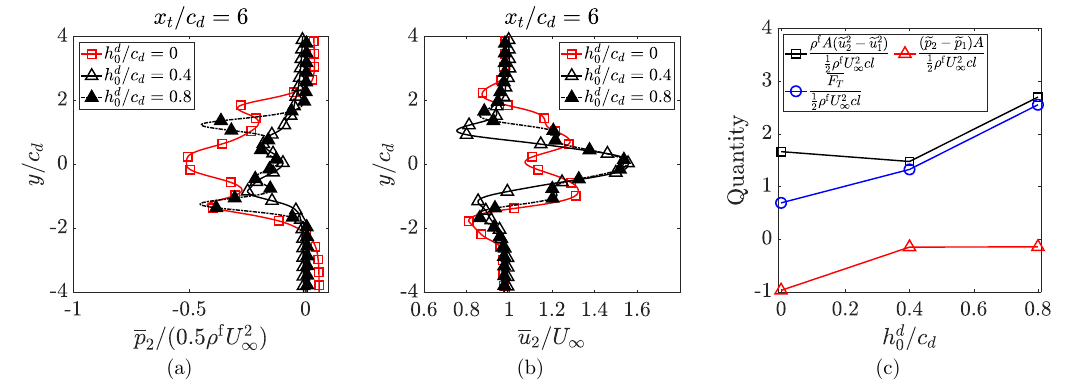}
        \caption{Control volume analysis of the tandem system of foils: (a) time-averaged pressure $\overline{p}_2$, (b) time-averaged streamwise velocity $\overline{u}_2$, at the right boundary of the control volume; and (c) evaluated quantities based on Eq. (\ref{Mom_Bal_Eq}) for representative $h_0^d/c_d$; for the gap ratio of $g/c_d = 7$.}
        \label{g7_pAvg_UAvg_h0d}
\end{figure}

Carrying out the control volume analysis similar to previous sections, the variation in the mean thrust force can be explained. The quantities $\widetilde{p}_2$ and $\widetilde{u}_2$ at the right boundary of the control volume are evaluated using Eq. (\ref{Mom_Bal_Eq}) with the help of time-averaged pressure $(\overline{p}_2)$ and velocity $\overline{u}_2$ which are plotted in Figs. \ref{g4_pAvg_UAvg_h0d} and \ref{g7_pAvg_UAvg_h0d} for $g/c_d = 4$ and 7, respectively. For both the gap ratios, the pressure distribution is such that the evaluated quantity $(\widetilde{p}_2 - \widetilde{p}_1)A$ does not change much for $h_0^d/c_d \in [0.4, 0.8]$ indicating the streamwise velocity distribution to be the dominant term representing the generated thrust. On the other hand, for $h_0^d/c_d \in [0, 0.4]$, the change in the velocity term $\rho^\mathrm{f}A(\widetilde{u}_2^2 - \widetilde{u}_1^2)$ is observed to be very small indicating the pressure term to be the dominant term and dictating the thrust trend. This observation is similar to that of the single isolated foil in Section \ref{single_h0_effects}.

\begin{figure}
		\centering
		\includegraphics[width=\textwidth]{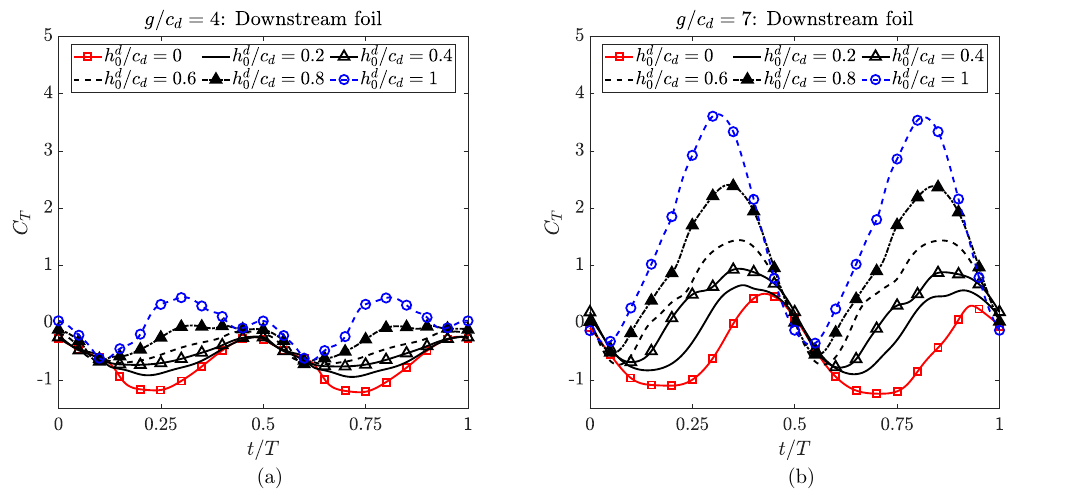}
        \caption{Temporal variation of the thrust coefficient for the downstream foil for varying $h_0^d/c_d$ at $g/c_d$: (a) 4, and (b) 7.}
        \label{TH_CT_h0d}
\end{figure}

\begin{figure}
		\centering
		\includegraphics[width=\textwidth]{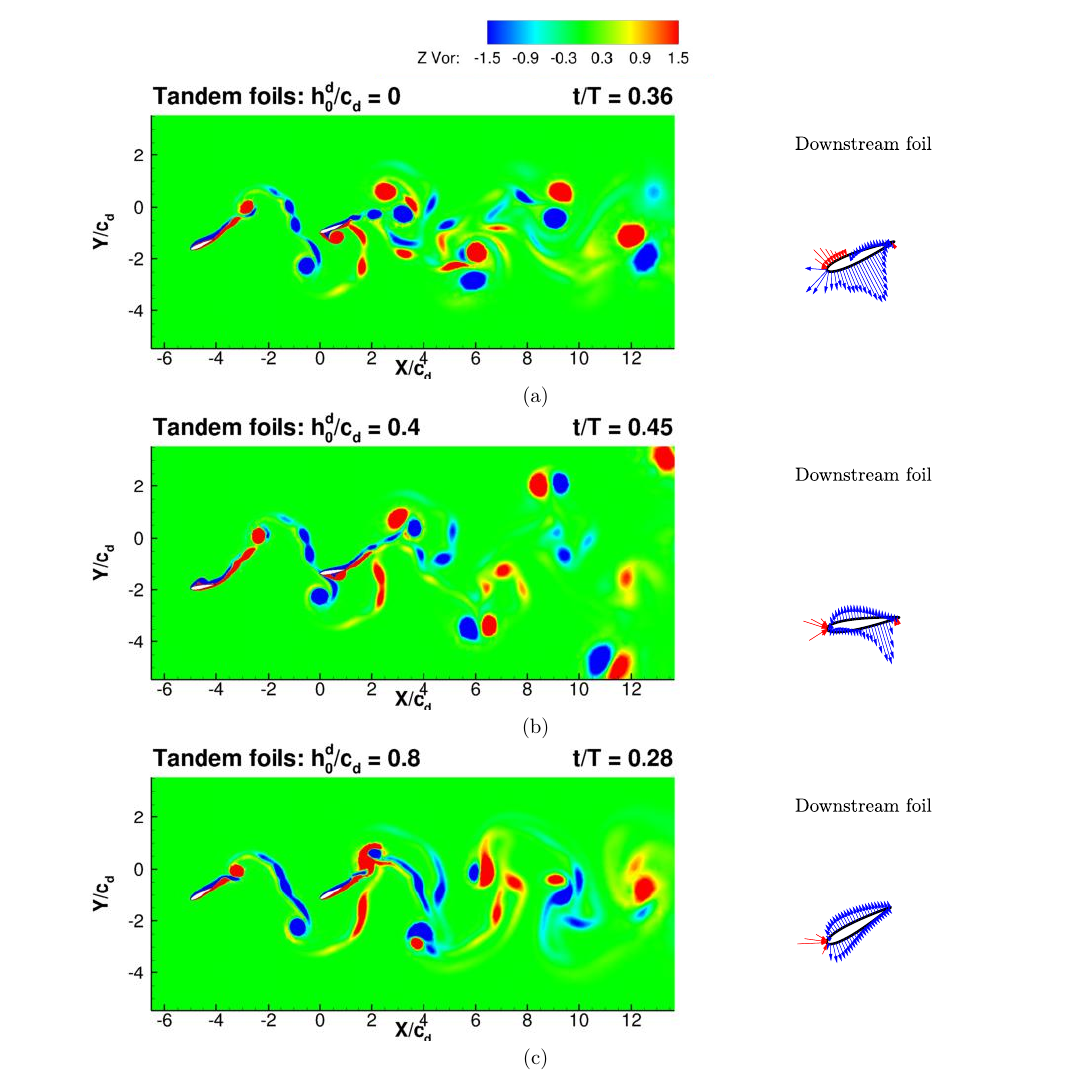}
        \caption{Flapping foils at different time instances of a flapping cycle at $g/c_d = 4$, $h^u_0/c_d = 1$, $\theta_0^u = \theta_0^d = 30^{\circ}$ and $h_0^d/c_d$: (a) 0, (b) 0.4, and (c) 0.8. The Z-vorticity contours and pressure distribution along the surface of the downstream foil are depicted on the Left and Right column at the instant.}
        \label{ZVor_tandem_g4_h0d}
\end{figure}

\begin{figure}
		\centering
		\includegraphics[width=\textwidth]{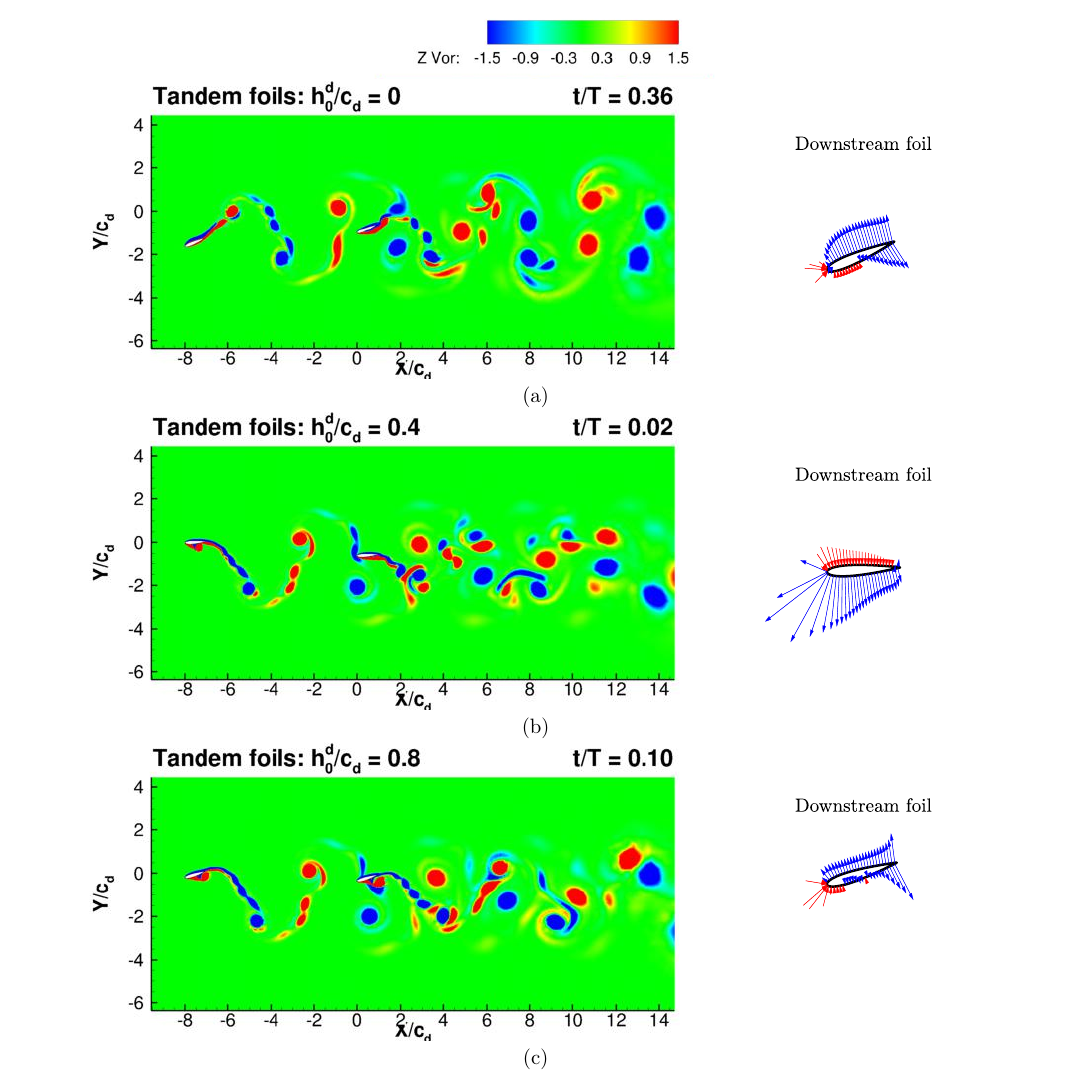}
        \caption{Flapping foils at different time instances of a flapping cycle at $g/c_d = 7$, $h^u_0/c_d = 1$, $\theta_0^u = \theta_0^d = 30^{\circ}$ and $h_0^d/c_d$: (a) 0, (b) 0.4, and (c) 0.8. The Z-vorticity contours and pressure distribution along the surface of the downstream foil are depicted on the Left and Right column at the instant.}
        \label{ZVor_tandem_g7_h0d}
\end{figure}

The temporal change in the thrust coefficient for the downstream foil is shown in Fig. \ref{TH_CT_h0d} for the two gap ratios. For both the gap ratios, the peak of the thrust coefficient in a flapping cycle increases with the heave amplitude $h_0^d/c_d$. However, this increase is significant for $g/c_d = 7$. We focus now on the vortex interaction of the upstream foil's wake with the downstream foil to get an insight about the instantaneous change in the thrust.
The vortex interaction can be visualized in Figs. \ref{ZVor_tandem_g4_h0d} and \ref{ZVor_tandem_g7_h0d} for $g/c_d = 4$ and $g/c_d = 7$, respectively. In Section \ref{single_h0_effects}, it was observed that for a single foil, at lower $h_0/c$, the wake is drag producing vK vortex street and as the heave amplitude increases, the propulsive performance increases with the wake resembling IvK vortex street. A similar observation can be inferred for the tandem configuration when the heave amplitude of the downstream foil is increased. For gap ratio of $g/c_d = 4$, the individual wake of the downstream foil is inherently vK vortex street at $h_0^d/c_d = 0$ (Fig. \ref{ZVor_tandem_g4_h0d}(a)). The incoming IvK wake of the upstream foil interacts with the downstream foil in such a manner that the opposite-signed vortices of the vK and IvK vortex streets pair up. This pairing can also be observed for $h_0^d/c_d = 0.4$ (Fig. \ref{ZVor_tandem_g4_h0d}(b)). With increase in $h_0^d/c_d$, the CCW vortex of the upstream foil's wake interacts with the downstream foil at the early of the downstroke resulting in I-2 where the CW vortex shear layer on the upper surface of the foil is prematurely shed (seen at $t/T = 0.28$ in Fig. \ref{ZVor_tandem_g4_h0d}(c)). Thus, for all the cases at $g/c_d = 4$, the thrust is negative (unfavorable condition) for the downstream foil and it is inherently unfavorable as a result of vK street for lower $h_0^d/c_d$. The negative thrust (or drag) for the downstream foil can also be noticed from the pressure arrow plots where most of the foil has suction pressure during the downstroke (unfavorable).
% Fig 16
% Fig 17

For higher gap ratio of $g/c_d = 7$, the vK vortex street for the downstream foil is observed at $h_0^d/c_d = 0$ (Fig. \ref{ZVor_tandem_g7_h0d}(a)) which leads to the negative thrust. With increase in the heave amplitude of the downstream foil, favorable condition for thrust generation occurs (Figs. \ref{ZVor_tandem_g7_h0d}(b) and (c)). The CW vortex from the wake of the upstream foil interacts with the upper and lower surface of the downstream foil. This results in supply of vorticity to the upper surface (I-1) and the vorticity on the lower surface is driven away (I-4). A pressure differential is created based on the suction pressure on the upper surface and high pressure on the lower surface of the foil during downstroke, leading to thrust. These interactions (I-1 and I-4) occur at the early stages of the downstroke as $h_0^d/c_d$ increases, thus extending the favorable condition in the duration where projected area of the foil to the freestream direction is maximum, giving higher average thrust.

\subsection{Effect of pitch amplitude of upstream foil $(\theta_0^u)$ on propulsion}

In this subsection, we examine the effect of the pitch amplitude of the upstream foil on the thrust and propulsive efficiency of the tandem configuration. Once again, two gap ratios ($g/c_d = 4,7$) are considered while the downstream pitch amplitude is fixed at $\theta_0^d = 30^\circ$. Here, the heave amplitudes of both the upstream and the downstream foil are held constant ($h_0^u/c_d = h_0^d/c_d = 1$).

\begin{figure}
		\centering
		\includegraphics[width=\textwidth]{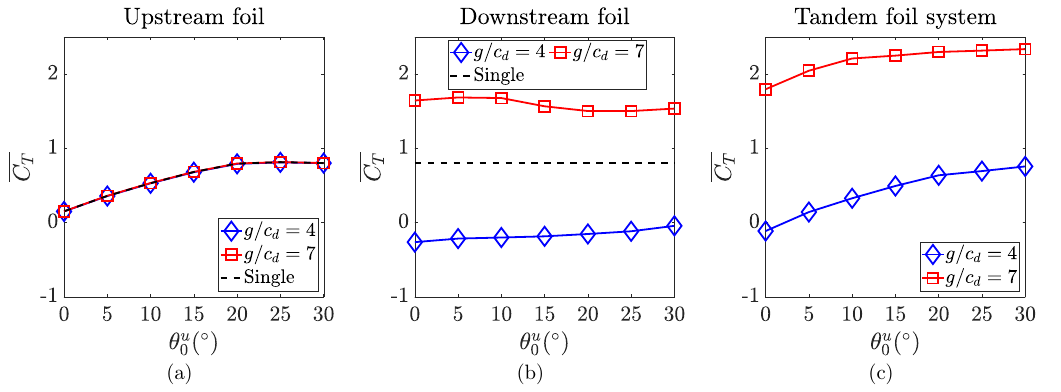}
        \caption{Variation of the mean thrust coefficient for the tandem flapping foils at $h^u_0/c_d = h^d_0/c_d = 1$, $\theta_0^d = 30^{\circ}$ with varying upstream foil pitch amplitude $\theta^u_0$ for (a) upstream foil, (b) downstream foil, and (c) tandem foil system.}
        \label{CTmean_theta0u}
\end{figure}

The mean thrust coefficient of the upstream foil for both the gap ratios (Fig. \ref{CTmean_theta0u}(a)) exhibits a similar behavior to that of the single foil indicating that the downstream foil does not influence the flow around the upstream foil. As the pitch amplitude ($\theta_0^u$) is increased to $30^\circ$ from zero, the mean thrust coefficient grows larger, peaking around $\theta_0^u = 25^\circ$ before dropping slightly at $\theta_0^u = 30^{\circ}$. The trend in $\overline{C_T}$ with $\theta_0^u$ can be explained in terms of the behavior of the single foil, explained in the previous section. The variation in the mean thrust coefficient for the downstream foil with $\theta_0^u$ is shown in Fig. \ref{CTmean_theta0u}(b). For $g/c_d = 4$, the thrust generated by the downstream foil increases slightly with pitch amplitude. On the other hand, at $g/c_d = 7$, the thrust generated by the downstream foil exhibits a more complex behavior; it decreases gradually until the region $\theta_0^u \in [20^\circ, 25^\circ]$ before starting to rise again. The thrust coefficient for a single foil at identical parameters ($h_0/c = 1$ and $\theta_0 = 30^{\circ}$) is also shown in the figure for comparison. Owing to the small variation in the mean thrust coefficient in this case, we can infer that the pitch amplitude of the upstream foil has a minor effect on the propulsive performance of the downstream foil. Therefore, for the combined tandem system, the mean thrust slightly increases as a result of the increasing trend for the upstream foil, as shown in Fig. \ref{CTmean_theta0u}(c), reaching a maximum of 2.34 and 0.76 for $g/c_d = 7$ and $4$, respectively, both at $\theta_0^u = 30^{\circ}$.
%Note that the mean thrust coefficient of an isolated foil does not depend on the pitch amplitude.
% Fig 18

\begin{figure}
		\centering
		\includegraphics[width=\textwidth]{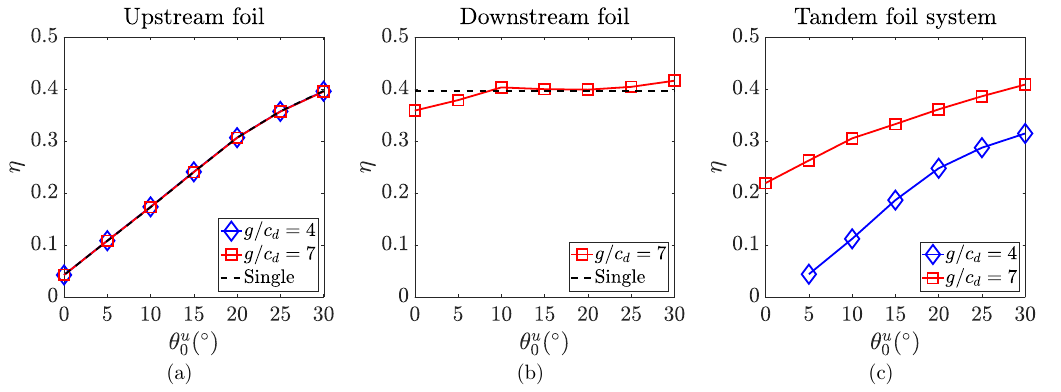}
        \caption{Variation of the propulsive efficiency for the tandem flapping foils at $h^u_0/c_d = h^d_0/c_d = 1$, $\theta_0^d = 30^{\circ}$ with varying upstream foil pitch amplitude $\theta^u_0$ for (a) upstream foil, (b) downstream foil, and (c) tandem foil system.}
        \label{Eta_theta0u}
\end{figure}

The propulsive efficiency of the upstream foil increases almost linearly with upstream pitch amplitude and negligible variation is observed as the gap ratio is changed, as shown in Fig. \ref{Eta_theta0u}(a). The trends for both $g/c_d = 4$ and $g/c_d = 7$ are identical to that of the single foil, as expected. On the other hand, the variation in the efficiency of the downstream foil is irregular for $g/c_d = 7$ (Fig. \ref{Eta_theta0u}(b)) as a slight variation between 0.36 and 0.415 is observed. Note that as the average thrust coefficient is negative for all the $g/c_d = 4$ cases, they are not shown in the efficiency plot. The propulsive efficiency for the combined tandem foils increases with $\theta_0^u$ for both the gap ratios reaching a maximum of 41\% and 31.4\% for $g/c_d = 7$ and $g/c_d = 4$, respectively, as observed in Fig. \ref{Eta_theta0u}(c). %Furthermore, it is observed that at $g/c_d = 7$ and $\theta_0^u = 0^{\circ}$, the force response of the downstream foil was chaotic and did not reach a periodic variation. 
% Fig 19

\begin{figure}
		\centering
		\includegraphics[width=\textwidth]{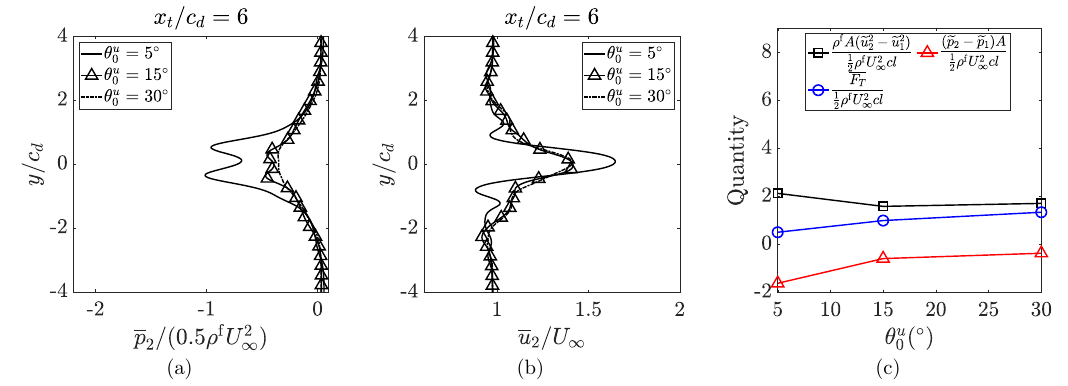}
        \caption{Control volume analysis of the tandem system of foils: (a) time-averaged pressure $\overline{p}_2$, (b) time-averaged streamwise velocity $\overline{u}_2$, at the right boundary of the control volume; and (c) evaluated quantities based on Eq. (\ref{Mom_Bal_Eq}) for representative $\theta_0^u$; for the gap ratio of $g/c_d = 4$.}
        \label{g4_pAvg_UAvg_theta0u}
\end{figure}

\begin{figure}
		\centering
		\includegraphics[width=\textwidth]{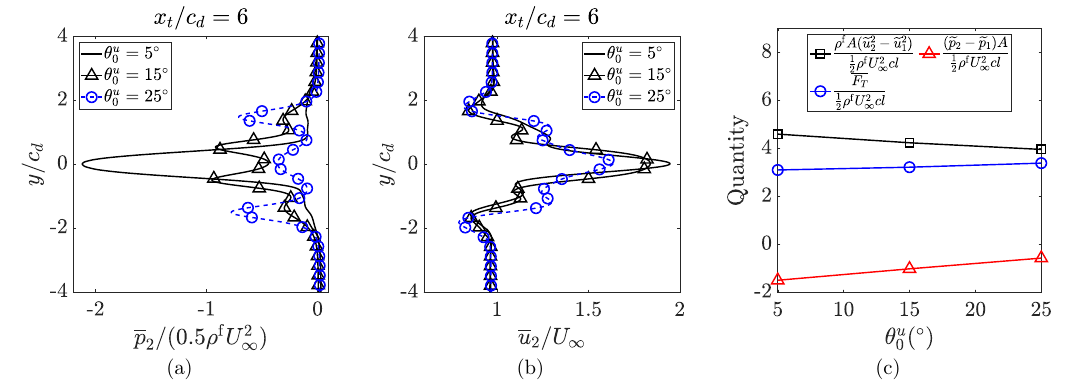}
        \caption{Control volume analysis of the tandem system of foils: (a) time-averaged pressure $\overline{p}_2$, (b) time-averaged streamwise velocity $\overline{u}_2$, at the right boundary of the control volume; and (c) evaluated quantities based on Eq. (\ref{Mom_Bal_Eq}) for representative $\theta_0^u$; for the gap ratio of $g/c_d = 7$.}
        \label{g7_pAvg_UAvg_theta0u}
\end{figure}

The control volume analysis provides an insight into the dominant term in determining the mean thrust generation. The pressure and streamwise velocity distribution on the right boundary of the control volume are shown in Figs. \ref{g4_pAvg_UAvg_theta0u} and \ref{g7_pAvg_UAvg_theta0u} for the two gap ratios. From the plots, it can be inferred that for $g/c_d = 4$, the term $(\widetilde{p}_2 - \widetilde{p}_1)A$ has a significant contribution for $\theta_0^u \in [5^{\circ}, 15^{\circ}]$ and its variation translates to the thrust generation. On the other hand, both the velocity and pressure terms tend to reach an asymptotic value as $\theta_0^u$ reaches $30^{\circ}$ leading to a minor increase in the mean thrust. At the gap ratio of 7, a significant difference is observed for the pressure distribution in Fig. \ref{g7_pAvg_UAvg_theta0u}(a) compared to the velocity (Fig. \ref{g7_pAvg_UAvg_theta0u}(b)) for various $\theta_0^u$ values. This gives an indication of the pressure term $(\widetilde{p}_2 - \widetilde{p}_1)A$ being a dominant factor in determining the thrust of the tandem foil system. Therefore, the mean thrust follows the trend of the pressure term and increases with $\theta_0^u$, albeit slowly.

\begin{figure}
		\centering
		\includegraphics[width=\textwidth]{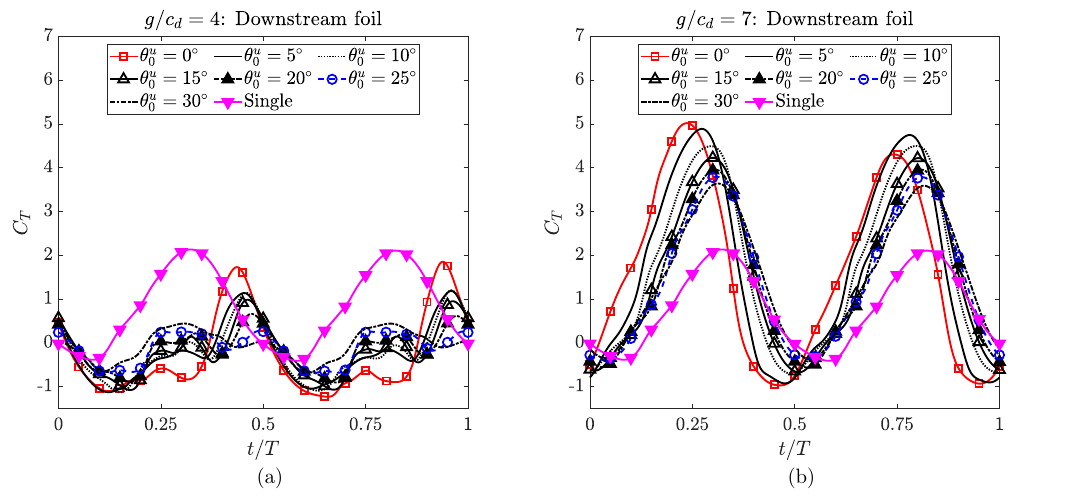}
        \caption{Temporal variation of the thrust coefficient for the downstream foil for varying $\theta_0^u$ at $g/c_d$: (a) 4, and (b) 7.}
        \label{TH_CT_theta0u}
\end{figure}

\begin{figure}
		\centering
		\includegraphics[width=\textwidth]{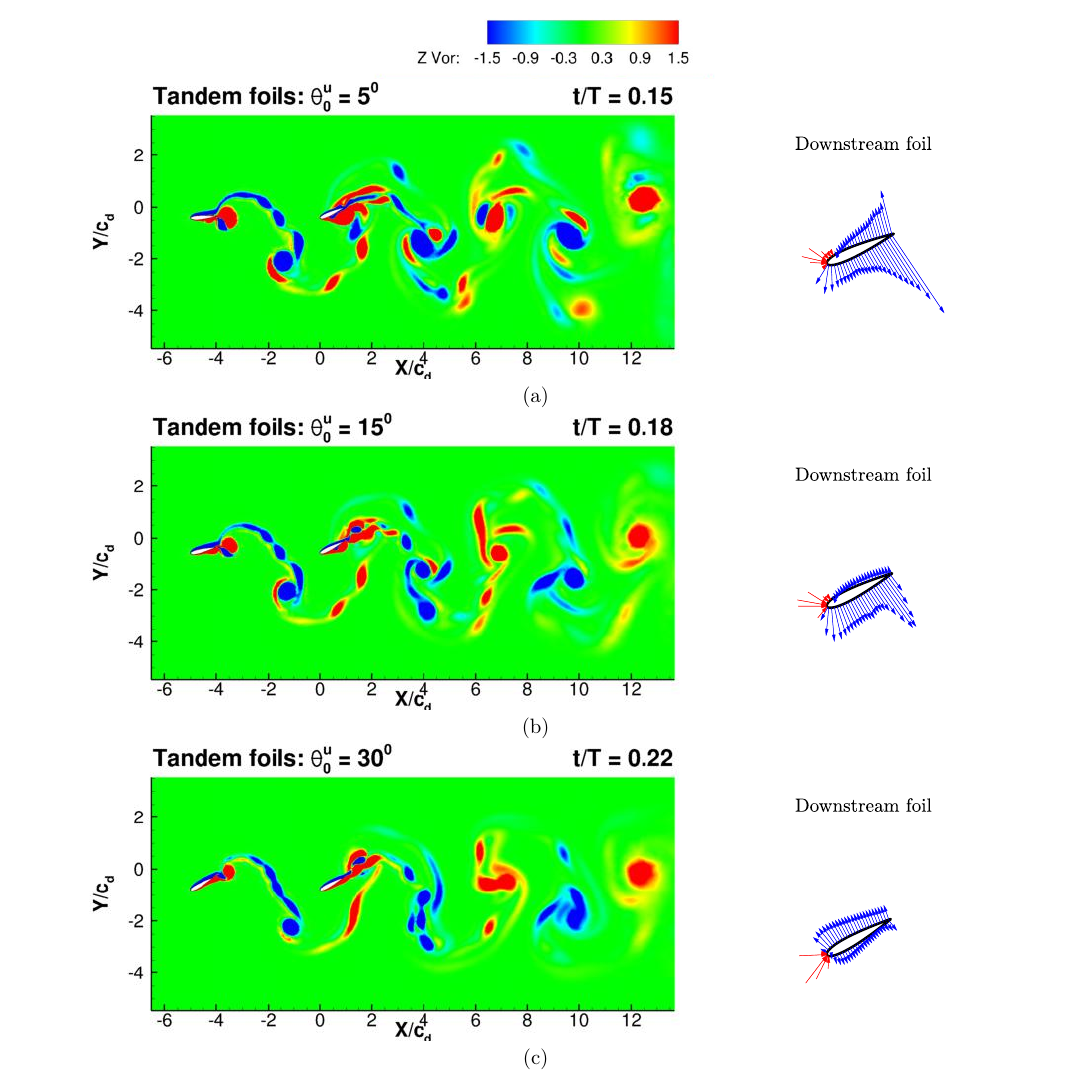}
        \caption{Flapping foils at different time instances of a flapping cycle at $g/c_d = 4$, $h^u_0/c_d = h^d_0/c_d = 1$, $\theta_0^d = 30^{\circ}$ and $\theta_0^u$: (a) $5^{\circ}$, (b) $15^{\circ}$, and (c) $30^{\circ}$. The Z-vorticity contours and pressure distribution along the surface of the downstream foil are depicted on the Left and Right column at the instant.}
        \label{ZVor_tandem_g4_theta0u}
\end{figure}

\begin{figure}
		\centering
		\includegraphics[width=\textwidth]{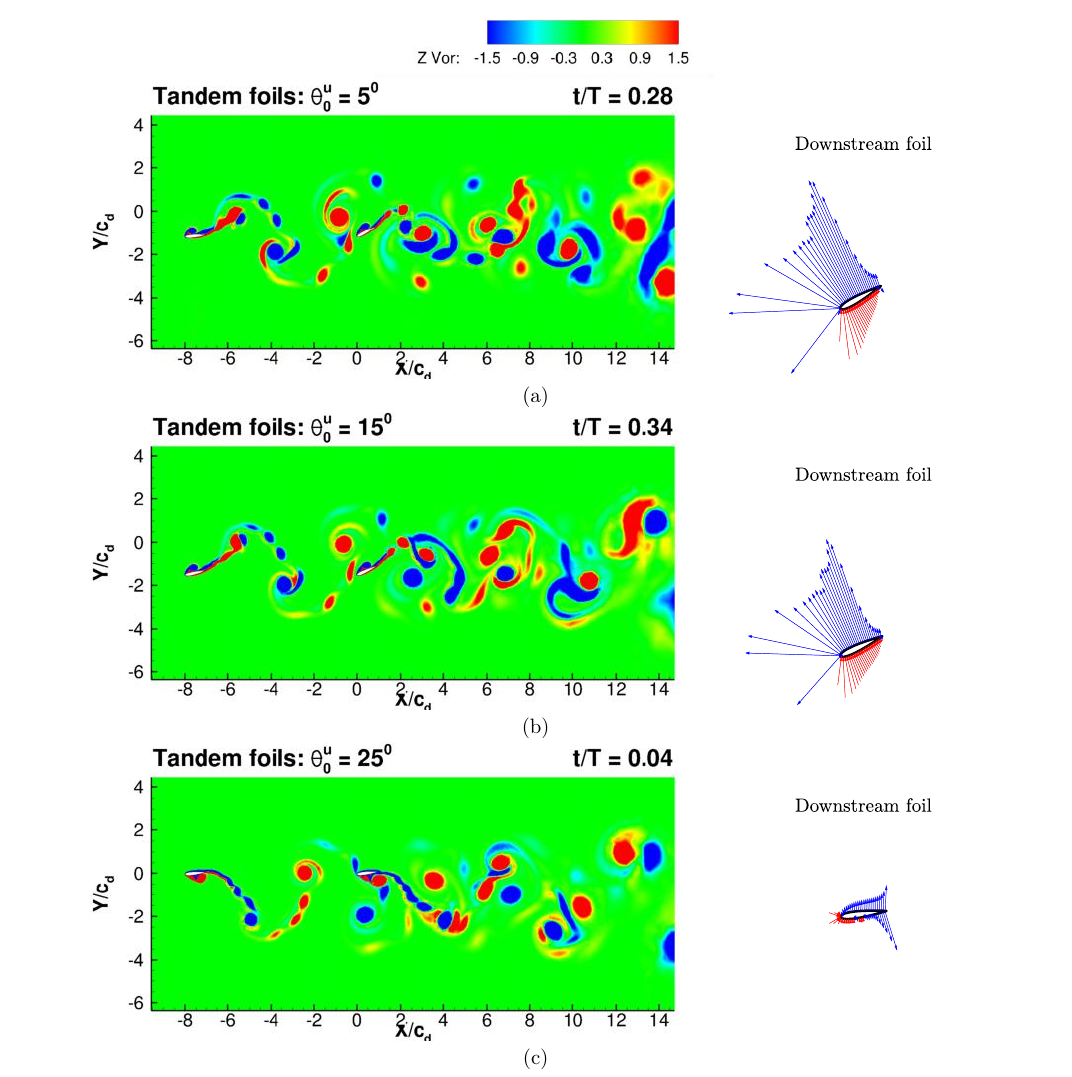}
        \caption{Flapping foils at different time instances of a flapping cycle at $g/c_d = 7$, $h^u_0/c_d = h^d_0/c_d = 1$, $\theta_0^d = 30^{\circ}$ and $\theta_0^u$: (a) $5^{\circ}$, (b) $15^{\circ}$, and (c) $25^{\circ}$. The Z-vorticity contours and pressure distribution along the surface of the downstream foil are depicted on the Left and Right column at the instant.}
        \label{ZVor_tandem_g7_theta0u}
\end{figure}

The instantaneous change in the thrust coefficient for the downstream foil at various $\theta_0^u$ is depicted in Fig. \ref{TH_CT_theta0u}. Minor changes are observed in the variation with the thrust coefficient at the mid-downstroke $(t/T = 0.25)$ in a flapping cycle; increasing and decreasing with increase in $\theta_0^u$ for $g/c_d = 4$ and $7$, respectively.
To elucidate the variation of the thrust of the downstream foil, we turn our attention to the Z-vorticity contours presented in Figs. \ref{ZVor_tandem_g4_theta0u} and \ref{ZVor_tandem_g7_theta0u} for $g/c_d = 4$ and $7$, respectively. It can be observed that as $\theta_0^u$ increases, the wake of the upstream foil is wider, identical to the observation for a single foil in Section \ref{single_theta0_effects}. For $g/c_d = 4$, the thrust coefficient is negative for all the cases as a consequence of unfavorable condition. However, there is a slight increase in the mean thrust with $\theta_0^u$. This can be explained by observing the increase in the delay of the shedding of the premature LEV by the CCW vortex as $\theta_0^u$ increases (as a consequence of the wider wake). The shedding occurs around $t/T = 0.15$, $0.18$ and $0.22$ for $\theta_0^u = 5^{\circ}$, $15^{\circ}$ and $30^{\circ}$, respectively (Fig. \ref{ZVor_tandem_g4_theta0u}). As the interaction gets delayed in the downstroke, the average thrust coefficient increases, although slightly. For higher gap ratio of $g/c_d = 7$, we observe a slight dip in $\overline{C_T}$ till $\theta_0^u = 20^{\circ}$ and then an increase till $30^{\circ}$. At $\theta_0^u = 5^{\circ}$ (Fig. \ref{ZVor_tandem_g7_theta0u}(a)), the stronger CCW vorticity from the upstream foil's wake is in close proximity to the downstream foil at $t/T = 0.28$ (around mid-downstroke). This leads to the favorable condition (I-3) which pulls the shear layer on the leading edge of the downstream foil, forming LEV. As the angle is increased to $\theta_0^u = 15^{\circ}$ in Fig. \ref{ZVor_tandem_g7_theta0u}(b), the interaction I-3 is weaker (due to the weaker CCW vortex) and delayed to $t/T = 0.34$. Furthermore, the stronger CCW vortex is far away from the upper surface due to the wider wake of the upstream foil at higher $\theta_0^u$. This delay in the interaction leads to a slightly lower mean thrust compared to $5^{\circ}$ and is the reason for the decreasing trend in $\overline{C_T}$. For $\theta_0^u = 25^{\circ}$, the favorable condition for thrust generation occurs via I-1 and I-4 during the start of the downstroke (Fig. \ref{ZVor_tandem_g7_theta0u}(c)) which results in increasing trend for the average thrust.  
% Fig 20
% Fig 21

\subsection{Effect of pitch amplitude of downstream foil $(\theta_0^d)$ on propulsion}

Having studied the effect of the pitch amplitude of the upstream foil, we now focus on the pitch amplitude of the downstream foil. The heave amplitudes of the upstream and downstream foils are fixed ($h_0^u/c_d = h_0^d/c_d = 1$) and the upstream pitch amplitude is $\theta_0^u = 30^\circ$ in this situation. 

\begin{figure}
		\centering
		\includegraphics{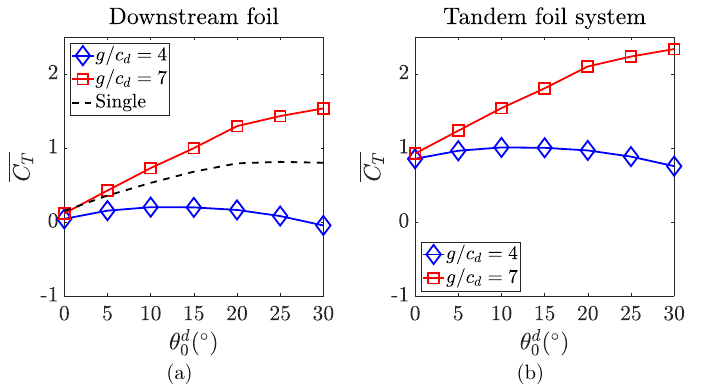}
        \caption{Variation of the mean thrust coefficient for the (a) downstream foil, and (b) tandem foil system, at $h^u_0/c_d = h^d_0/c_d = 1$, $\theta_0^u = 30^{\circ}$ with varying downstream foil pitch amplitude $\theta^d_0$.}
        \label{CTmean_theta0d}
\end{figure}

The influence of $\theta_0^d$ on the mean thrust coefficient of the downstream foil is shown in Fig. \ref{CTmean_theta0d}(a). For the reference case of an isolated (single) foil, the mean thrust coefficient increases until $\theta_0^d = 25^\circ$ after which it gradually settles at $\overline{C_T} \approx 0.8$. In comparison, for $g/c_d = 4$, $\overline{C_T}$ increases and then decreases with $\theta_0^d$. The peak value of $\overline{C_T}$ occurs at $\theta_0^d = 10^\circ$ and it is noted that $\overline{C_T}$ for all the cases is very close to zero. 
%More specifically, the mean thrust coefficient for the case of $g/c_d = 4$ actually becomes negative at $\theta_0^d = 30^\circ$ indicating that the feathering limit has been crossed. 
On the other hand, for $g/c_d = 7$, the mean thrust coefficient increases consistently for $\theta_0^d \in [0^\circ, 30^\circ]$ with a maximum value close to 1.6. At low values of $\theta_0^d$, the behavior of the mean thrust is observed to be similar to that of a single foil. The combined tandem system behaves similarly to the downstream foil characteristics as the thrust of the upstream foil (constant) gets added to that of the downstream foil.
% Fig 22

\begin{figure}
		\centering
		\includegraphics{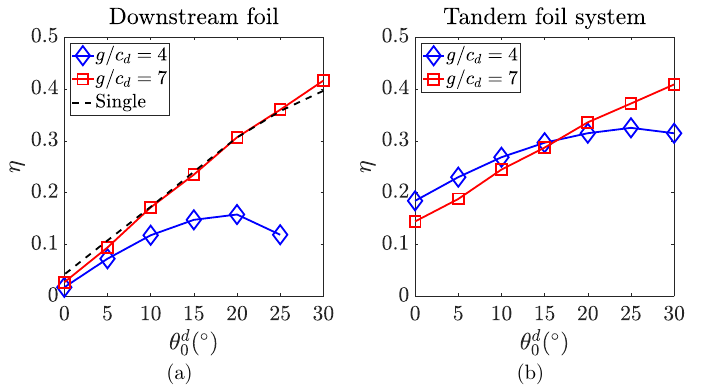}
        \caption{Variation of the propulsive efficiency for the (a) downstream foil, and (b) tandem foil system, at $h^u_0/c_d = h^d_0/c_d = 1$, $\theta_0^u = 30^{\circ}$ with varying downstream foil pitch amplitude $\theta^d_0$.}
        \label{Eta_theta0d}
\end{figure}

The variation in propulsive efficiency is shown in Fig. \ref{Eta_theta0d}. Here, the isolated foil exhibits an almost linear increase in $\eta$ with $\theta_0^d$. For $g/c_d = 4$, the propulsive efficiency of the downstream foil increases with downstream pitch amplitude before peaking at $\theta_0^d = 20^\circ$ and then decreasing again. When the gap ratio is increased to $g/c_d = 7$, the behavior of $\eta$ closely matches that of the isolated foil case. On the other hand, it is interesting to note that the efficiency of the combined system increases monotonically for both the gap ratios, as shown in Fig. \ref{Eta_theta0d}(b).
% Fig 23

\begin{figure}
		\centering
		\includegraphics[width=\textwidth]{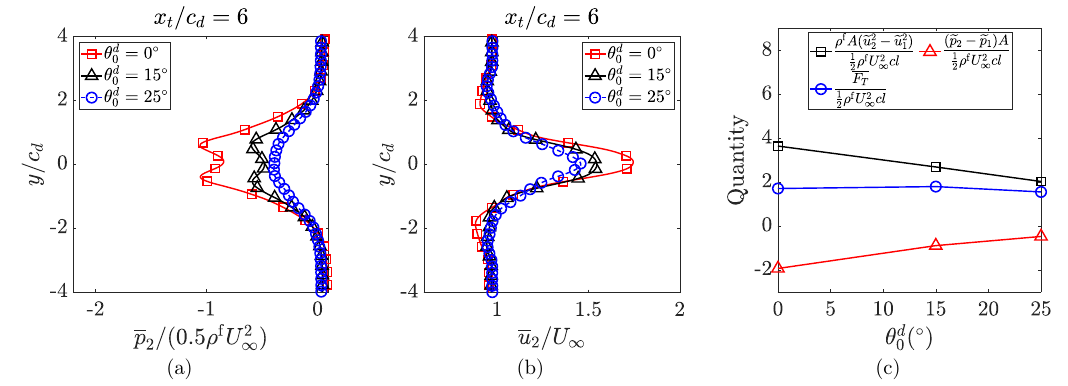}
        \caption{Control volume analysis of the tandem system of foils: (a) time-averaged pressure $\overline{p}_2$, (b) time-averaged streamwise velocity $\overline{u}_2$, at the right boundary of the control volume; and (c) evaluated quantities based on Eq. (\ref{Mom_Bal_Eq}) for representative $\theta_0^d$; for the gap ratio of $g/c_d = 4$.}
        \label{g4_pAvg_UAvg_theta0d}
\end{figure}

\begin{figure}
		\centering
		\includegraphics[width=\textwidth]{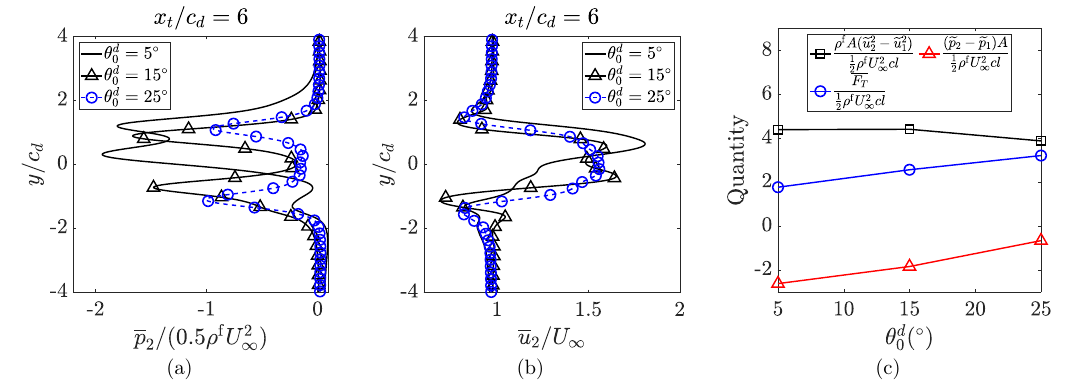}
        \caption{Control volume analysis of the tandem system of foils: (a) time-averaged pressure $\overline{p}_2$, (b) time-averaged streamwise velocity $\overline{u}_2$, at the right boundary of the control volume; and (c) evaluated quantities based on Eq. (\ref{Mom_Bal_Eq}) for representative $\theta_0^d$; for the gap ratio of $g/c_d = 7$.}
        \label{g7_pAvg_UAvg_theta0d}
\end{figure}

The control volume analysis shows a similar variation as that of a single foil presented in Section \ref{single_theta0_effects}. The time averaged pressure and velocity for the two gaps are provided in Figs. \ref{g4_pAvg_UAvg_theta0d} and \ref{g7_pAvg_UAvg_theta0d}. For $g/c_d = 4$, the combined effects of the velocity and pressure distributions drive the thrust variation with $\theta_0^d$. The thrust increases slightly till $\theta_0^d = 15^{\circ}$ and then decreases with further increase in the pitch amplitude. The competing effects of the two terms leads to a minor variation of the thrust with $\theta_0^d$. At high gap ratio of 7, the contribution of the velocity term $\rho^\mathrm{f}A(\widetilde{u}_2^2 - \widetilde{u}_1^2)$ seems to be small across $\theta_0^d$ (see Fig. \ref{g7_pAvg_UAvg_theta0d}(c)). Thus, the pressure term dictates the trend in the thrust coefficient, resulting in an increase in $\overline{C_T}$ with $\theta_0^d$.

\begin{figure}
		\centering
		\includegraphics[width=\textwidth]{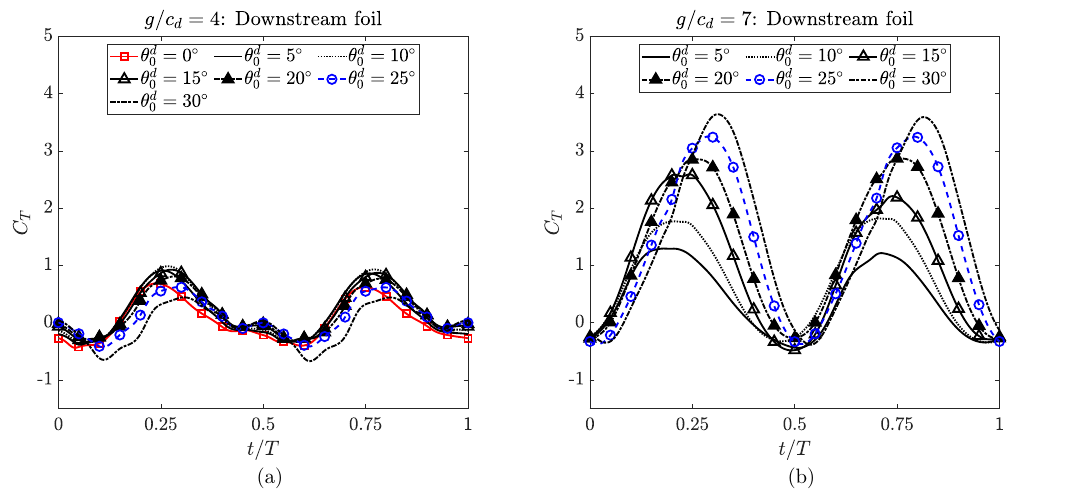}
        \caption{Temporal variation of the thrust coefficient for the downstream foil for varying $\theta_0^d$ at $g/c_d$: (a) 4, and (b) 7.}
        \label{TH_CT_theta0d}
\end{figure}

\begin{figure}
		\centering
		\includegraphics[width=\textwidth]{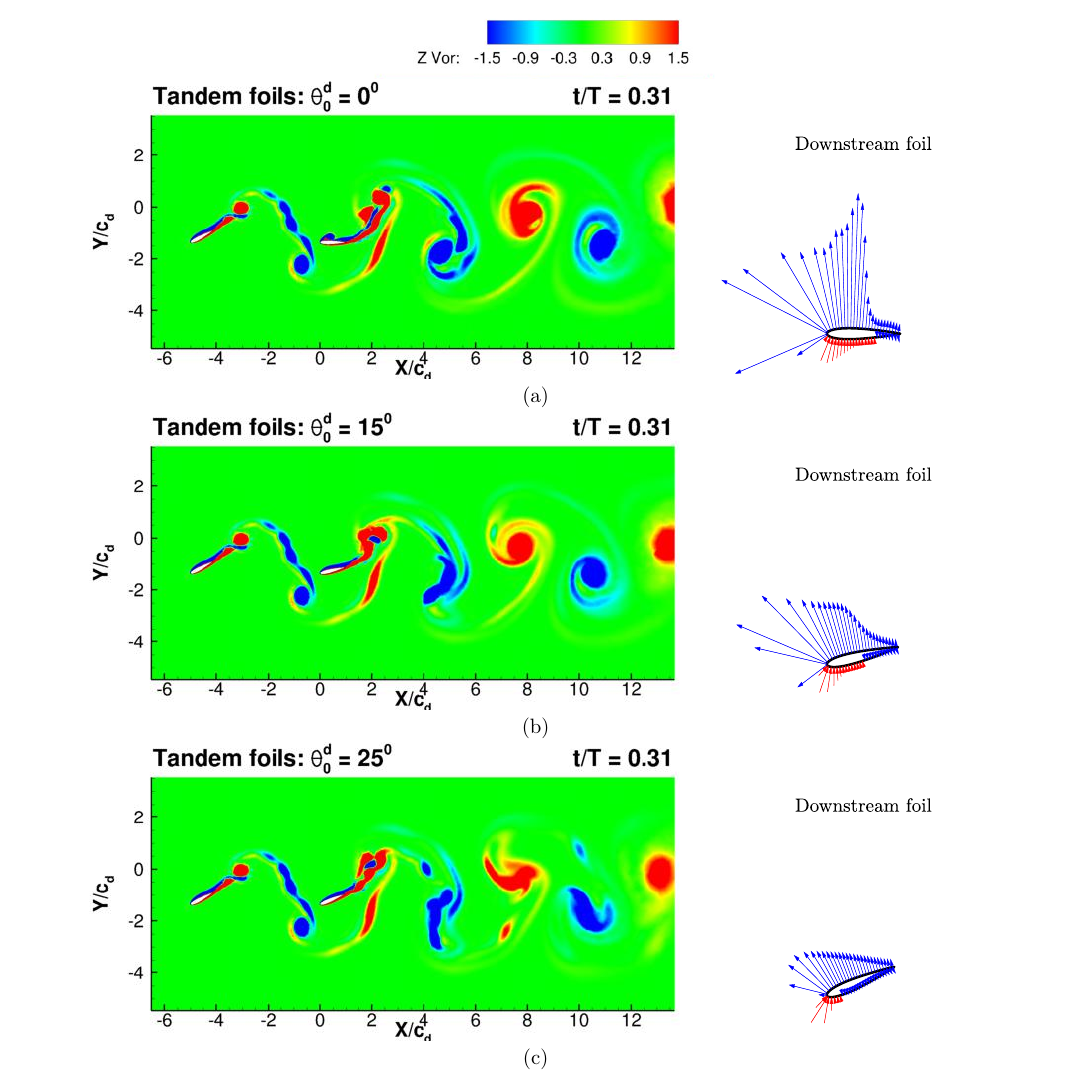}
        \caption{Flapping foils at different time instances of a flapping cycle at $g/c_d = 4$, $h^u_0/c_d = h^d_0/c_d = 1$, $\theta_0^u = 30^{\circ}$ and $\theta_0^d$: (a) $0^{\circ}$, (b) $15^{\circ}$, and (c) $25^{\circ}$. The Z-vorticity contours and pressure distribution along the surface of the downstream foil are depicted on the Left and Right column at the instant.}
        \label{ZVor_tandem_g4_theta0d}
\end{figure}

The time history of the thrust coefficient for the downstream foil is shown in Fig. \ref{TH_CT_theta0d} for the two gap ratios. A negligible variation for various $\theta_0^d$ for $g/c_d = 4$ is observed. However, the maximum thrust in a cycle increases with $\theta_0^d$ for the higher gap ratio of 7.
Let us further analyse the flow dynamics to comprehend the variation in the propulsive performance. As observed for a single isolated foil, the mean thrust coefficient increases with $\theta_0$ for $f^* = 0.2$ as the frontal area to the freestream velocity $U_{\infty}$ increases. Therefore, if the upstream foil would not be present, one would observe an increase in $\overline{C_T}$. In the case of tandem configuration, the behavior of the downstream foil is influenced by the interaction of it with the wake of the upstream foil, leading to either favorable or unfavorable conditions for propulsion. For $g/c_d = 4$, there seems to be a competing effect between the increasing frontal area to freestream velocity of the downstream foil and the effect of the unfavorable condition. For all the cases of $\theta_0^d$, there is an unfavorable condition caused due to the interaction of CCW vortex of the upstream foil's wake with the leading edge of the downstream foil (I-2 and I-5) resulting in premature LEV shedding. For $\theta_0^d = 0^{\circ}$ (Fig. \ref{ZVor_tandem_g4_theta0d}(a)), a secondary LEV is formed during the downstroke. As $\theta_0^d$ increases, the projected frontal area of the downstream foil also increases leading to an increase in $\overline{C_T}$ at $\theta_0^d = 15^{\circ}$ (Fig. \ref{ZVor_tandem_g4_theta0d}(b)). However with further increase in $\theta_0^d$, there is no generation of secondary LEV and premature shedding of LEV is more dominant leading to decrease in average thrust (Fig. \ref{ZVor_tandem_g4_theta0d}(c)). Therefore, at lower $\theta_0^d$, the downstream foil behaves similar to an isolated foil (increase in mean $C_T$ with $\theta_0^d$: although slightly due to already existing unfavorable condition), while for higher $\theta_0^d$, the unfavorable condition is more prominent and reduces the propulsive performance of the downstream foil. 
% Fig 24

\begin{figure}
		\centering
		\includegraphics[width=\textwidth]{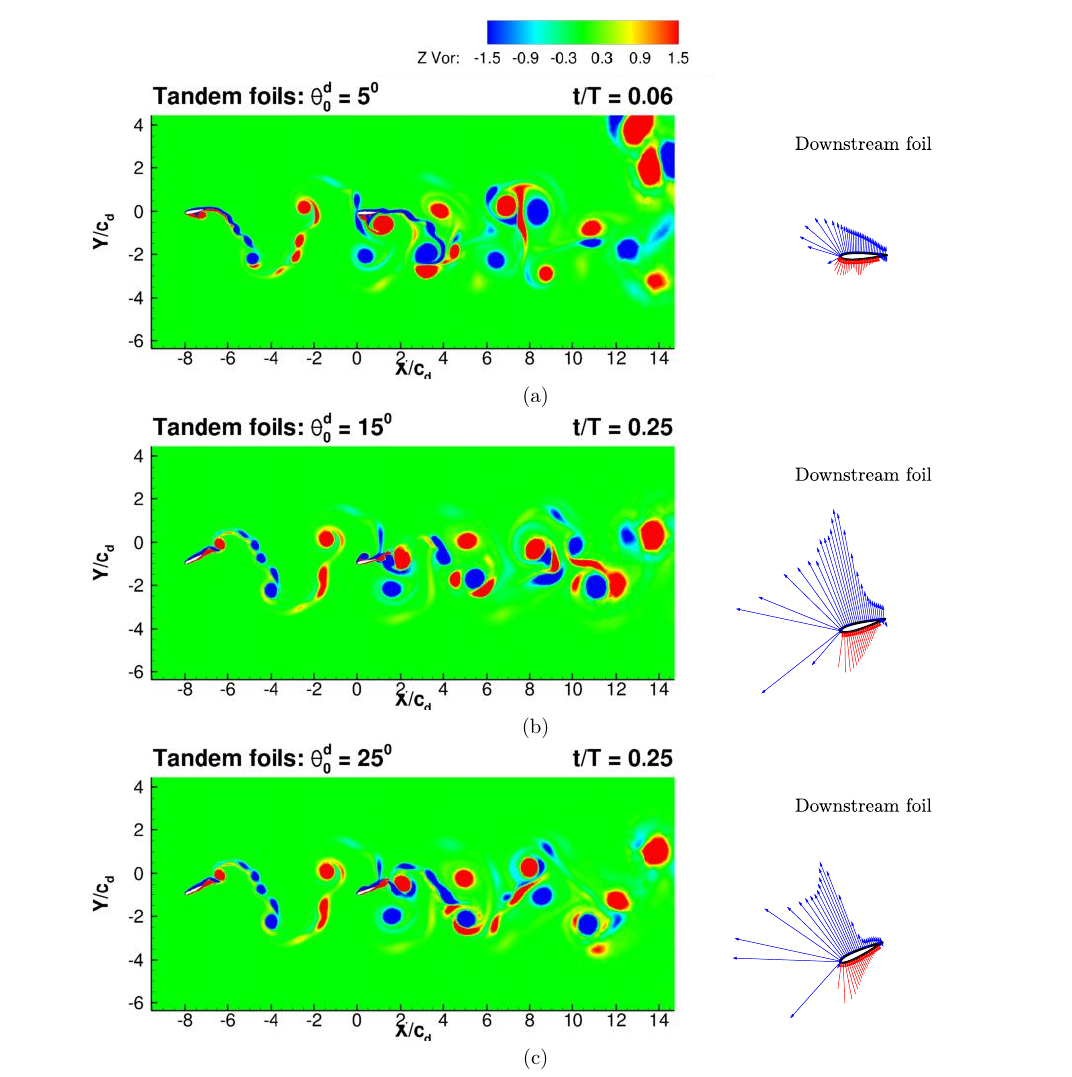}
        \caption{Flapping foils at different time instances of a flapping cycle at $g/c_d = 7$, $h^u_0/c_d = h^d_0/c_d = 1$, $\theta_0^u = 30^{\circ}$ and $\theta_0^d$: (a) $5^{\circ}$, (b) $15^{\circ}$, and (c) $25^{\circ}$. The Z-vorticity contours and pressure distribution along the surface of the downstream foil are depicted on the Left and Right column at the instant.}
        \label{ZVor_tandem_g7_theta0d}
\end{figure}

Contrary to low gap ratio, we observe an increase in mean thrust for $g/c_d = 7$. The favorable thrust generating condition as a consequence of I-1 and I-4 can be observed for $\theta_0^d = 5^{\circ}$ in Fig. \ref{ZVor_tandem_g7_theta0d}(a). A comparison of the vorticity contours at the quarter time period for $\theta_0^d = 15^{\circ}$ and $25^{\circ}$ is depicted in Fig. \ref{ZVor_tandem_g7_theta0d}(b-c). The favorable condition occurs for all the $\theta_0^d$ cases as indicated by the pressure arrow plots showing negative and positive pressures on the upper and lower surfaces of the foil, respectively. However, the projected frontal area to the freestream velocity increases with $\theta_0^d$ leading to an increase in the component of the net force in the freestream direction, i.e., thrust. 
% Fig 25

Therefore, the type of vortex interaction from the wake of the upstream foil with the downstream foil; and the time and duration of interaction in the flapping cycle determines the propulsive performance of the downstream foil. From an operational perspective, the tandem foil system generates maximum thrust where interactions leading to favorable conditions occur at the downstream foil. Moreover, the propulsion efficiency is determined by the amount of thrust generated and the power input to the system. Optimal regimes where the thrust is higher and the power input is lower leads to higher efficiency of the propulsive system.

\section{Three-dimensional tandem flapping foils}
\label{3d_demo}
\begin{figure}
	\centering
		    \includegraphics[width=\textwidth]{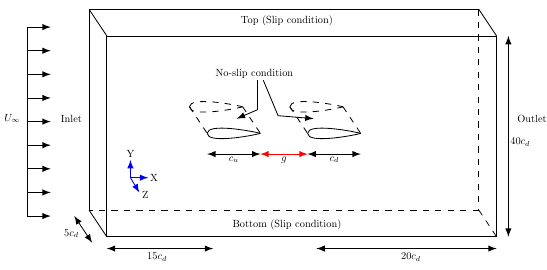}
\caption{Schematic of the three-dimensional computational domain for a uniform flow across tandem flapping foils at $Re=1100$.}
	\label{3D_foil_schematic}
\end{figure}

In order to understand if the wake-foil interaction in tandem flapping exhibits any three-dimensionality, we investigate the tandem foil system considering the parameters 
%As a demonstration of the present numerical formulation and its scalability, 
%we perform a numerical computation considering the three-dimensional domain of the tandem flapping foils. 
%The parameters selected are 
$Re = 1100$, $g/c_d = 7$, $h_0^u/c_d = h_0^d/c_d = 1$, $\theta_0^u = \theta_0^d = 30^{\circ}$, $\phi^u_h = \phi^d_h = 90^{\circ}$ and $\varphi = 0^{\circ}$. The computational domain with the boundary conditions is depicted in Fig. \ref{3D_foil_schematic}. A freestream velocity in the X-direction is imposed on the inlet boundary which is $15c_d$ from the upstream foil, while a stress-free condition is satisfied at the outlet boundary ($20c_d$ from the downstream foil). The slip or no-penetration condition is satisfied at the top and bottom boundaries which are equidistant ($20c_d$) from the center of the foils. A no-slip condition is satisfied at the surface of the two foils. The two-dimensional mesh created for the computations presented in the previous sections is extruded in the third dimension with a span size of $5c_d$ considering $\Delta z/c_d = 0.125$ as the resolution in the Z-direction. The boundaries perpendicular to the Z-axis representing the total span of the foils are considered to be periodic boundaries. Note that this configuration models the foil to have an infinite span, eliminating the end-effects. The discretization consists of around 4.1 million grid points consisting of approximately 4 million eight-node hexahedral elements. 
% Fig 32

\begin{figure}
		\centering
		\includegraphics[width=\textwidth]{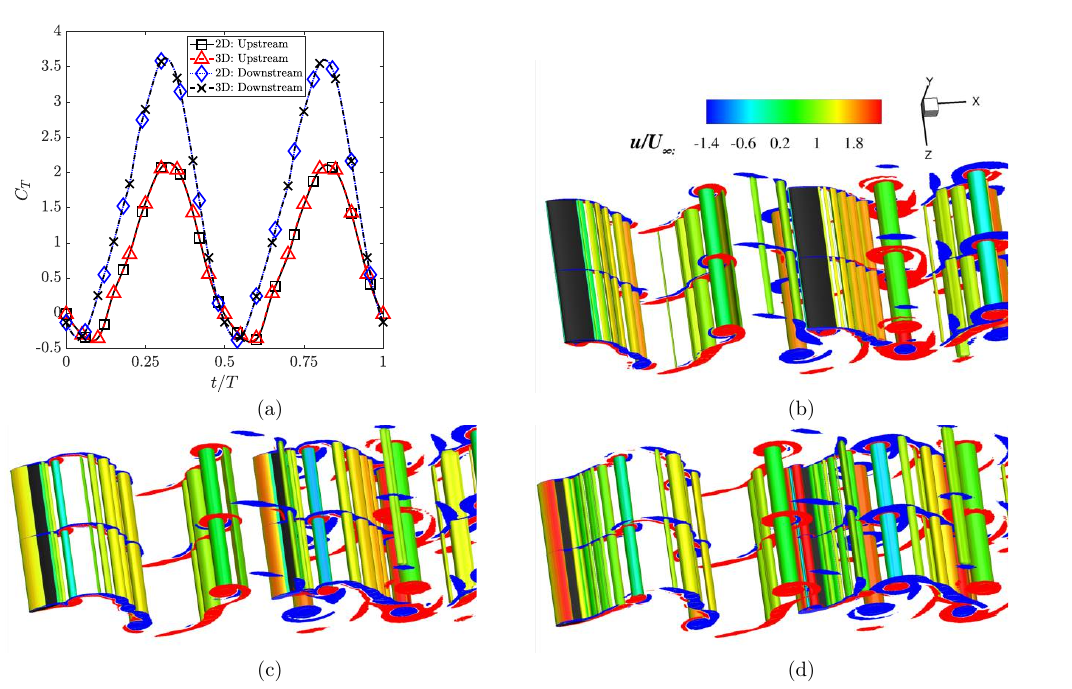}
        \caption{Three-dimensional flapping at $g/c_d = 7$, $h^u_0/c_d = h^d_0/c_d = 1$, $\theta_0^u = \theta_0^d = 30^{\circ}$, $f^*_u = f^*_d = 0.2$ and $Re = 1100$: (a) comparison of the thrust coefficient in a flapping cycle for 2D and 3D computations, and iso-surfaces of $Q$-criterion colored by streamwise velocity (and Z-vorticity at various layers along the span) at $t/T$: (b) 0, (c) 0.2, and (d) 0.4.}
        \label{3d_flapping}
\end{figure}

The time history of the thrust coefficient for the upstream and the downstream foils is shown in Fig. \ref{3d_flapping}(a). The temporal variation of the thrust coefficient for the two-dimensional case has also been plotted. It can be observed that for the parameters considered, there is no difference between the propulsive response of the tandem configuration of flapping foils in three-dimensions as compared to the two-dimensional results. This is further corroborated by the visualization of the vortex structures in Fig. \ref{3d_flapping}(b-d). The three-dimensional vortex structures are depicted by the iso-surfaces of Q-criterion at 0.25 colored by the freestream velocity. The two-dimensional Z-vorticity is also shown at three different spans of $z/c_d \in [0, 2.5, 5]$. In spite of the large amplitude flapping of the foils and the wake-foil interaction in the tandem configuration, the flow structures are observed to be two-dimensional at $Re = 1100$. This suggests the adoption of the two-dimensional simulation can be employed for understanding the flow dynamics at this Reynolds number. %It can be concluded that no three-dimensionality of flow is observed at $Re = 1100$ for the tandem flapping foils, suggesting the adoption of the two-dimensional simulations to be correct. 
Further study needs to be carried out at high Reynolds numbers including turbulence effects to understand the three-dimensionality of the flow patterns.
% Fig 33

\section{Conclusions}
\label{conclusion}

The present study numerically investigated the influence of heave $(h_0^u/c_d \in [0, 1]$ and $h_0^d/c_d \in [0, 1])$ and pitch amplitudes $(\theta_0^u \in [0^{\circ}, 30^{\circ}]$ and $\theta_0^d \in [0^{\circ}, 30^{\circ}])$ on the propulsion of a single and tandem foil system. The combination of these kinematic parameters was explored and their effects on the performance of NACA 0015 foils in tandem were studied. %The two-dimensional variational formulation has been extended to three-dimensions to demonstrate the three-dimensional flow effects due to flapping at low Reynolds number for tandem foils.

%\redcolor{Insights about the propulsive performance of the foils can be gained by comprehending the wake interaction mechanisms with the foil.} 
For a single isolated foil, the lift and the drag forces were observed to be directly proportional with the effective angle of attack and the projected area of the foil to the incoming flow direction, respectively. Through a control volume analysis, it was shown that both the time-averaged pressure and the streamwise velocity at the wake are crucial indicators to quantify the mean thrust force. Corroborating the findings from the literature \citep{vanBuren2019, Yu2017}, the propulsive performance of the single foil increased with heave amplitude pertaining to the increase in the effective angle of attack resulting in larger lift force, which has a major contribution to the thrust. Increase in pitch amplitude also increased the propulsive force as a consequence of increasing projected area of the foil to the freestream velocity resulting in higher component of the thrust force.  

The salient findings from the investigation of the tandem foil system at two gap ratios are summarized as follows:

$\bullet$ The upstream foil behaved as the single isolated foil with no interference from the downstream foil for the considered streamwise gaps between the foils.

$\bullet$ The time-averaged pressure and streamwise velocity variation in the wake of the tandem system collectively determined the mean propulsive force for the system, as shown by the control volume analysis. The trend of the mean thrust for the foil system is similar to that of single foil when considering the effects of heave $(h_0^d)$ and pitch $(\theta_0^d)$ amplitudes of the downstream foil.

$\bullet$ The transient behavior of the thrust for the tandem system can be interpreted by examining the vortex interaction patterns and linking it with favorable and unfavorable conditions for thrust generation provided in the literature \citep{Joshi2021, Broering2012, gopalkrishnan_triantafyllou_triantafyllou_barrett_1994, Lewin_2003, muscutt_weymouth_ganapathisubramani_2017}.

$\bullet$ At the low Reynolds number of $1100$, the wake structures due to the wake-foil interaction are observed to be two-dimensional for large amplitude combined heave and pitch motions.
%in the tandem foil system 
%generated by the combined heave $h^u_0 = h^d_0$ and pitch motions of the tandem system is observed to be two-dimensional. 

%\section*{Data Availability}
%The data that supports the findings of this study are available within the article.

\section*{Acknowledgments}
The computational work for this article was performed on the High Performance Computing resource at BITS Pilani, K K Birla Goa Campus and the resource at Computing Lab, Department of Mechanical Engineering, BITS Pilani, Hyderabad Campus.

\section*{Declaration of Interests}
The authors report no conflict of interest.

\bibliographystyle{jfm}
\bibliography{references}
	
\end{document}